\documentclass[12pt,aps,ams,amsfonts,nofootinbib]{revtex4}

\newtheorem{lema}{Lemma}
\newtheorem{cor}{Corollary}
\newtheorem{propo}{Proposition}
\newtheorem{teor}{Theorem}

\begin{document}

\title{Spectral conditions on the state of a composite quantum system implying its separability}

\author{G.A. Raggio}
\email{raggio@famaf.unc.edu.ar}
\affiliation{FaMAF-UNC, C\'ordoba, Argentina}

\date{May 4, 2005}

\begin{abstract}
 The separability modulus $\ell ( \rho)$ of a state $\rho$ of an arbitrary finite composite quantum system, is the largest $t$ in $[0,1]$ such that $t.\rho +(1-t).\tau$ is unentangled, where $\tau$ is the normalized trace. The basic properties of $\ell$, introduced by Vidal \& Tarrach \cite{ViTa} in another guise, are briefly established. With these properties, we obtain conditions on the spectrum of a state which imply that it is separable. As a consequence, we show that for any Hamiltonian $H$  the thermal equilibrium states $e^{- H/T}/tr(e^{-H/T})$ are  separable  if  $T$ is large enough.
 Also, for $F$ a unitarily invariant, convex  continuous real-valued function on states, for which $F(\rho )> F(\tau )$ whenever $\rho \neq \tau$,  there is a critical $C_F$ such that $F( \rho )\leq C_F $ implies that $\rho$ is separable, and for each possible $c > C_F$ there are entangled states $\phi$ with $F( \phi )=c$. This class includes all strictly convex  unitarily invariant continuous functions, and also every non-trivial partial eigenvalue-sum. Some $C_F$'s are computed. General upper and lower bounds for $C_F$ are given,  and then improved for bipartite systems.
\end{abstract}
\maketitle

\newcommand{\CC}{\mathbb C}
\newcommand{\RR}{\mathbb R}

\newcommand{\dem}{\noindent \underline{Proof}: }

\section{Introduction}

Consider a finite level quantum system described by a complex  Hilbert space ${\cal H}$ of finite dimension $d$. Let ${\cal B} ( {\cal H})$ be the linear operators from ${\cal H} $ into itself equipped with the operator norm. An operator $ a \in {\cal B}({\cal H})$ is termed positive, written $a \geq 0$,  if $\langle \psi , a \psi \rangle \geq 0$ for all $\psi \in {\cal H}$. An application of  polarization identity  shows that $a\geq 0$ implies that $a$ is self-adjoint. A state $\rho$ on ${\cal B}({\cal H})$ is a linear function $\rho : {\cal B}({\cal H})\to {\mathbb C}$ which is positive, that is $a \geq 0$ implies $\rho (a) \geq 0$, and normalized $\rho ( {\bf 1})=1$, where ${\bf 1}$ is the identity operator.  It follows that $\rho (a^*)=\overline{\rho (a)}$. The state space ${\cal S} ( {\cal H})$ or simply ${\cal S}$ is the set of all states. ${\cal S}$ is a convex subset of the set of linear functionals on ${\cal B}({\cal H})$ which is compact with respect to the topology defined by the norm of linear functionals $f$ given by $\parallel f \parallel = \sup \{|f(a)|/\parallel a \parallel : \;  0 \neq a \in {\cal B}({\cal H})\}$. All continuity statements made in the present paper refer to this topology. \\

The extremal points $ext ({\cal S})$ are precisely the pure states or vectorial states given by $\rho (a) = \langle \psi , a \psi \rangle$ for some unit vector $\psi \in {\cal H}$. We reserve the term ``pure'' for these states. The decomposition of a non-pure $\rho \in {\cal S}$ as a convex sum $\rho= \sum_{j=1}^M t_j \rho^{(j)}$, with $t_j >0$, $\sum_{j=1}^M t_j =1$ and the $\rho^{(j)}$'s are pure (where $M$ can be $\infty$ in which case the convex sum always converges in norm no matter what the states $\rho^{(j)}$ are) is never unique; there are always uncountably many such decompositions with finite $M$. Among these convex decompositions into pure states, the {\em spectral decompositions} are those for which the pure states $\rho^{(j)}$ involved are pairwise orthogonal meaning that the associated vectors $\{\psi_j\}$ are pairwise orthogonal. For a spectral decomposition one has $M \leq d$ and the spectral decomposition is unique iff all the weights (non zero eigenvalues of the density operator, see below)  $t_j>0$ involved are distinct.\\

Now there is a well-known (and in finite dimension elementary) representation theorem for states which says that every $\rho \in {\cal S}$ is given by $ \rho (a) = tr (D_{\rho}a)$ where $D_{\rho}$ is a unique density operator; that is $D_{\rho} \geq 0$ and $tr (D_{\rho} )=1$. The density operators ${\cal B}_1^+ ( {\cal H})$ clearly form a convex set which turns out to be compact with respect to the trace norm on ${\cal B}({\cal H})$ given by $\parallel a\parallel_1 = tr ( |a| )$. The formulas $\rho (a) = tr (D_{\rho}a)$, $\rho \in {\cal S}$; and $\rho_D (a) = tr (Da)$, $D \in {\cal B}_1^+ ( {\cal H})$,  implement a biyective affine  homeomorphism between the compact convex sets ${\cal S}$ and ${\cal B}_1^+ ({\cal H})$. One has $ext ({\cal B}_1^+ ({\cal H}))= \{ p\in {\cal B}( {\cal H}): \; p=p^*=p^2\;,\; rank (p)=1\}$, i.e. the orthoprojectors onto the one-dimensional subspaces of ${\cal H}$. The spectral theorem applied to $D_{\rho}$ provides a spectral decomposition of $\rho$. In what follows we often identify $\rho$ with its density operator. \\

Given $N \geq 2$ finite dimensional Hilbert spaces ${\cal H}_j$ of dimension $d_j\geq 2$ ($j=1,2, \cdots , N$), the composite quantum system --whose constituents are the quantum systems described by ${\cal H}_j$-- is described by the tensor-product ${\cal H}={\cal H}_1\otimes {\cal H}_2 \otimes \cdots \otimes {\cal H}_N$ which has dimension $D=d_1d_2\cdots d_N$. Given a state $\rho \in {\cal S}$ we can define a state of ${\cal B}({\cal H}_j)$ by
\[ \rho^{[j]} (a) = \rho (\underbrace{{\bf 1}_1\otimes {\bf 1}_2  \otimes \cdots \otimes {\bf 1}_{j-1}}_{\small \mbox{$j-1$ identity factors}} \otimes a \otimes \underbrace{{\bf 1}_{j+1} \otimes \cdots {\bf 1}_N}_{\small \mbox{$N-j$ identity factors}})\;,\; a \in {\cal B}({\cal H}_j)\;.\]
A state $\rho$ is  said to be a {\em product-state} if
\[ \rho (a_1\otimes a_2 \otimes \cdots \otimes a_N) = \rho^{[1]} ( a_1)\rho^{[2]}(a_2) \cdots  \rho^{[N]}(a_N)\;,\]
for all $a_1 \in {\cal B}({\cal H}_1)$, all $a_2 \in {\cal B}({\cal H}_2)$, $\cdots$, and all $a_N \in  {\cal B}({\cal H}_N)$. Clearly, a product-state exhibits no correlations whatsoever among the constituent subsystems. The product-states in ${\cal S}$ are denoted by ${\cal S}^{prod}$ and they are closed.\\

Given states $\rho_j $ of ${\cal B}({\cal H}_j)$, the map
\[ (\rho_1\otimes \rho_2\otimes \cdots \otimes \rho_N ) (a_1\otimes a_2\otimes \cdots a_N ) = \rho_1(a_1)\rho_2 (a_2) \cdots \rho_N (a_N )\; ,\]
$a_j \in {\cal B}({\cal H}_j)$, admits a unique extension (by linearity and continuity) to a state of ${\cal B}({\cal H})$ which is denoted by
$\rho_1\otimes \rho_2\otimes \cdots \otimes \rho_N$. Thus, $\rho\in {\cal S}^{prod}$  iff $\rho =\rho^{[1]}\otimes \rho^{[2]}\otimes \cdots \otimes \rho^{[N]}$.\\

We come to the basic definitions. Recall that the convex hull $co ({\cal K})$ of a subset ${\cal K}$ of a convex set is the collection of all finite convex sums of elements of ${\cal K}$. A state $\rho \in {\cal S}$ is said to be {\em separable} if it lies in the convex hull $co ({\cal S}^{prod})$ of the product-states. We write ${\cal S}^{sep}$ for the separable states. Due to the finite-dimension, ${\cal S}^{sep}=co ({\cal S}^{prod}) $ is a closed and thus compact subset of ${\cal S}$.
The extremal points of ${\cal S}^{sep}$, $ext ({\cal S}^{sep})$, are precisely the pure product-states $ext ({\cal S})\cap {\cal S}^{prod}$. Thus in analyzing the separability of a given $\rho \in {\cal S}$ one can restrict oneself to the convex decompositions of $\rho$ into pure states. As mentioned, there are uncountably many such finite decompositions and, in general, the spectral decomposition(s) of a separable state are not decompositions into product-states. This is what makes the problem of deciding whether a state is separable or not  a very subtle problem. A state is called {\em entangled} if it is not separable; that is if it cannot be decomposed into a convex sum of (pure) product-states.\\

At present there are finite algorithms deciding whether a given state is separable or not only for two qubits ($N=2$ with $d_1=d_2=2$; (Wootters Criterion, \cite{Wo}; PPT Criterion, \cite{HHH}) and for $N=2$ with $d_1=2$ and $d_2=3$ (PPT Criterion, \cite{HHH}). L.Gurvits \cite{Gu} has shown that the separability problem is NP-hard in the category of computational complexity theory \footnote{In this paper, we deal exclusively with the finite-dimensional case but: what happens when one or more of the Hilbert spaces involved is a separable infinite-dimensional Hilbert space? There are two alternatives. One can restrict oneself to the {\em normal} states ${\cal S}_*$ (those states that are continuos with respect to the weak-operator topology on ${\cal B}({\cal H})$, see \cite{Dix}) to proceed and then ${\cal S}_*^{sep}$ is defined to be the closure of $co( {\cal S}_*^{prod})$. Or, alterantively, one can consider all states and then it is natural to define ${\cal S}^{sep}$ to be the closure of $co ( {\cal S}^{prod})$ with respect to the weak toplogy on functionals induced by ${\cal B}({\cal H})$ to get a compact set. Be that as it may, there is almost no hope, today,  of dealing with the separability problem in either case.}. \\

 Here we present some very elementary arguments and basic facts which nevertheless allow us to isolate simple conditions on the spectrum of  a density operator of an arbitrary finite composite quantum system guaranteeing that the state is separable. One of the  basic ingredients of our arguments is not new: If $\tau$ is the normalized trace which is a product state (hence separable) and $\rho$ is any state, how large can $t$ get before $t.\rho +(1-t).\tau $ becomes entangled?
This question underlies the work of \.{Z}yczkowski et al, (\cite{Zy}), Vidal and Tarrach (\cite{ViTa}), and other authors.
Here we proceed backwards, in as much as we do not extend previous work, but use very elementary methods which require almost no precise information about entanglement to extract some general results, that seem to have been overlooked. We show that for any unitarily invariant, concave (or convex) real-valued function $F$ on states, for which $F(\rho )=F( \tau )$ implies $\rho =\tau$, there exists a critical value $C_F$ such that $F(\rho ) \geq C_F$ implies that $\rho$ is separable. The class of functions with this separating property includes all unitarily invariant strictly convex or concave continuous functions, and also the non-trivial partial eigenvalue-sums (which define the ``more mixed than'' partial ordering of states). Another simple result is that for any Hamiltonian of an arbitrary  composite quantum system, there are finite critical temperatures $T_c^+\geq0$ and $T_c^-\leq0$ such that  the thermal equilibrium state $\exp (-H/T)/tr(\exp ( -H/t))$ is separable
if  $T \geq T_c^+$ or $T\leq T_c^-$. With precise entanglement information, as is available or obtainable for $N=2$, the results of this paper can be considerably improved and sharp bounds can be obtained for the critical values mentioned. Some results in this direction are obtained here, but work on this is in progress.\\

The organization of the paper is as follows. Section II and its subsections, where the composite structure is irrelevant, presents some basic facts about global spectral properties of states and introduces a representation, the gap-representation, of a state which turns out to be  useful.  In section III, we turn to composite systems and, in \S III.A, introduce and present basic material about the separability modulus of a state (the critical $t$ of the above paragraph); this quantity has been studied extensively in another guise by Vidal and Tarrach \cite{ViTa}. In \S III.B we obtain some simple spectral conditions which are sufficient for separability. The rest of the subsections of \S III, deal with the mentioned application to thermal states and unitarily invariant convex  functions with the separation property. In \S IV, we use available information about the modulus of separability for bipartite systems to precise the results of \S III.  Appendix A, contains the only honest, yet trivial, ``entanglement calculation'' showing that if the state of a bipartite system has eigenvalues $0$ and $(D-1)^{-1}$, where $D$ is the dimension, then it has positive partial transpose. \\

\section{The global spectral properties of states}

In this section  we consider states for some fixed $d$-dimensional Hilbert  space ${\cal H}$ with $d\geq 2$ and abbreviate ${\cal B}_d ={\cal B}({\cal H})$ and  ${\cal S}_d ={\cal S}({\cal H})$. We write $\tau_d$ for the state given by the normalized trace $\tau_d (a) = tr(a)/d $, $a \in {\cal B}_d$.
We often identify the state with the density matrix associated to it.  For any selfadjoint matrix $A \in{\cal B}_d $  write $spec (A)=(a_1,a_2, \cdots , a_d)$ for the vector in ${\mathbb R}^d$ whose entries are the eigenvalues of $A$ taking into account their multiplicities and numbered nonincreasingly: $a_1\geq a_2 \geq \cdots \geq a_d$. If $d=6$ and $A$ has eigenvalues 17 with multiplicity 2; $\pi^2$ with multiplicity 1; and $0$ with multiplicity 3, then $spec (A)= (17,17, \pi^2, 0,0,0)$. \\

\subsection{The spectral simplex}\label{SS}

If $\rho$ is a state of ${\cal B}_d$ with $spec ( \rho ) = ( \lambda_1, \lambda_2, \cdots , \lambda_d )$ then, $1\geq \lambda_1 \geq \lambda_2 \geq \cdots \geq \lambda_d \geq 0$, and $\sum_{j=1}^d \lambda_j =1$.
We write $s_-(\rho )$ for the minimal eigenvalue of $\rho$; it satisfies $s_-(\rho ) \leq 1/d$ there being equality iff $\rho =\tau_d$.\\

We introduce the set of all possible spec's of states
\[ {\cal L}_d := \{ (\lambda_1, \lambda_2, \cdots , \lambda_d ): \; \lambda_1\geq \lambda_2 \geq \cdots \geq \lambda_d \geq 0 \;,\; \sum_{j=1}^d \lambda_j =1\}\; ,\]
and have that ${\cal L}_d$ is the image of ${\cal S}_d$ under the map $spec$. If $u \in {\cal B}_d$ is unitary and $\rho \in {\cal S}_d$, then $\rho_u$ defined by $\rho_u (a) = \rho ( u^* au)$, is a state; and $spec ( \rho_u ) = spec ( \rho )$. We say ${\cal K}\subseteq {\cal S}_d$ is unitarily invariant if $\rho \in {\cal K}$ implies $\rho_u\in {\cal K}$ for every unitary $u\in {\cal B}_d$. The map $spec : {\cal S}_d \to {\cal L}_d$ is continuous (use  singular value inequalities \cite{HJ}, or alternatively the second resolvent equation). The ordering required in the definition of the map $spec$ prevents it form being affine; for example if $\rho_1, \rho_2$ are pairwise orthogonal pure states then $spec( \rho_1)=spec (\rho_2)=(1,0,\cdots , 0)$, $spec ( t \cdot \rho_1+(1-t)\cdot \rho_2)=(\max\{t,1-t\},\min\{t,1-t\}, 0,\cdots , 0)$.\\

Let $\rho \in {\cal S}_d$ and put $\lambda=spec ( \rho )$. Take a spectral decomposition of $\rho$, $\rho = \sum_{j=1}^d \lambda_j \cdot \rho^{(j)}$ where $\{\rho^{(j)}:\; j=1,2,\cdots , d\}$ is a maximal family of pairwise orthogonal pure states associated with an orthonormal basis $\{\psi_j : \; j=1,2, \cdots , d\}$ of ${\cal H}$. Consider any of the $d$ cyclic permutations $\pi $ of length $d$  of $d$ objects, and let $\pi ( \rho )=\sum_{j=1}^d \lambda_{\pi (j)}\cdot \rho^{(j)}$. Then obviously $spec( \pi ( \rho))=\lambda$ and there is a unitary $u\in {\cal B}_d$ with $\pi ( \rho )=\rho_u$ (i.e., $u\psi_k=\psi_{\pi^{-1}(k)}$). Moreover, $\sum_{\pi } (1/d) \pi ( \rho )=\tau_d$. This proves
\begin{lema} \label{TAU} For any state $\rho$, $\tau_d = \sum_{j=1}^d (1/d) \pi_j (\rho)$ where the $\pi_j$'s are the cyclic permutations of length $d$ of $d$ objects. The states $\pi_j ( \rho )$ are unitarily equivalent to $\rho$ and have the same $spec$ as $\rho$.\\
\end{lema}

 The geometric structure of ${\cal L}_d$ is simple. It is a $(d-1)$-simplex, that is a convex set with $d$ extremal points such that the decomposition of every one of its points into a convex sum of extremal points is unique.

\begin{propo} \label{LD} ${\cal L}_d$ is a compact convex subset of ${\mathbb R}^d$ and a $(d-1)$-simplex. The  $d$ extremal points are given by the vectors
\[ e^{(k)}=(\underbrace{1/k,1/k,\cdots, 1/k}_{\mbox{$k$ times}},0, \cdots , 0)\;,\; k=1,2, \cdots , d \;.\]
If $\lambda = (\lambda_1,\lambda_2, \cdots , \lambda_d )\in {\cal L}_d$ then
\[ \lambda = \sum_{j=1}^{d} x_j e^{(j)} \;,\]
where
\[ x_j = j(\lambda_j - \lambda_{j+1}) \;,\;\; j=1,2, \cdots , d-1 \;, \; \; x_d = d \lambda_d \;,\]
and $\sum_{j=1}^d x_j =1$.\\

If $x_j \geq 0$ for $j=1,2, \cdots , d$ and $\sum_{j=1}^d x_j =1$, then
\[  \sum_{j=1}^d x_j e^{(j)} = \left( \sum_{k=1}^d x_k/k , \sum_{k=2}^d x_k/k, \cdots , \sum_{k=j}^d x_k/k , \cdots , x_d/d\right)       \;,\]
that is the $j$-th component of $\sum_{j=1}^d x_j e^{(j)}$ is $\sum_{k=j}^d x_k/k$.

\end{propo}

\dem Convexity and compactness are clear. We show that
\[ e^{(k)}=(\underbrace{1/k,1/k,\cdots, 1/k}_{\mbox{$k$ times}},0, \cdots , 0)\]
is extremal for each $k=1,2,\cdots,d$. Suppose $x,y \in {\cal L}_d$ and $0<t<1$. If
$e^{(k)}=tx+(1-t)y$ then
\[  x_j=y_j=0 \;,\;\mbox{ for $j=k+1,k+2, \cdots, d$}\]
and, thus
\[ tx_j+(1-t)y_j = 1/k \;,\;\mbox{ for $j=1,2,\cdots ,k$}\;,\]
\[  \mbox{and $x_1+x_2+\cdots +x_k =y_1+y_2+\cdots +y_k=1$}\;.
\]
But then, for $j=1,2,\cdots , k-1$ we have
\[ t(x_j-x_{j+1})+(1-t)(y_j-y_{j+1})=0 \;,\]
which implies $x_j=x_{j+1}$ and $y_j=y_{j+1}$ for all $j=1,2,\cdots , k-1$ and thus $x_j=y_j=1/k$
for all $j=1,2,\cdots , k$.
Hence $x=y=e^{(k)}$ proving that $e^{(k)}$ is extremal.
In order to prove that there are no other extremal points it suffices to show that every $\lambda \in {\cal L}_d$ is a convex combination of
these $d$ extremal points. Now,
\[ \sum_{j=1}^d x_j e^{(j)} = \left( \sum_{\ell =1}^d \frac{x_{\ell}}{\ell},  \sum_{\ell =2}^d \frac{x_{\ell}}{\ell} , \cdots, \sum_{\ell =k}^d \frac{x_{\ell}}{\ell}, \cdots, \frac{x_d}{d} \right) \;;\]
and the equation $\sum_{j=1}^d x_j e^{(j)}=\lambda$ for arbitrary $\lambda \in {\cal L}_d$ can be solved for
the $x_j$'s recursively giving
\[ x_d= d x_d \;,\;\; x_{j}=j \left( \lambda_j - \lambda_{j+1}\right) \;,
\;\; j=1,2,\cdots , d-1 \;,\]
which are unique. Clearly $x_j \geq 0$ for all $j=1,2,\cdots , d$ and
\[ \sum_{j=1}^d x_j = d \lambda_d + \sum_{j=1}^{d-1} j(\lambda_j-\lambda_{j+1})=
d \lambda_d +\sum_{j=1}^{d-1} j\lambda_j -\sum_{\ell =2}^{d} (\ell -1)\lambda_{\ell} \]
\[= d x_d +\sum_{j=1}^{d-1} (j-(j-1))\lambda_j -(d-1)\lambda_d =
\sum_{j=1}^d \lambda_j=1\;. \;\Box\\ \]

Recall the theory of majorization for vectors in ${\cal L}_d$ (\cite{AlUh},\cite{HJ}). The connections with entanglement are reviewed in \cite{NiVi}\footnote{We adhere to the time-honored convention of \cite{AlUh}, which is distinct from that of \cite{HJ}, which is different than that of \cite{NiVi}, which is not the same as that of \cite{AlUh}.}.
Define the $k$-th partial sum $\Sigma_k ( \lambda ) $ of $\lambda \in {\cal L}_d$ by $\Sigma_k ( \lambda )=\sum_{j=1}^k \lambda_j $ ($k=1,2, \cdots , d$), which is affine in $\lambda$. Agree that for $\lambda , \mu \in {\cal L}_d$, $\lambda \succ \mu $ means that $\Sigma_k ( \lambda )\leq \Sigma_k ( \mu )$ for every $k=1,2, \cdots , d$. We observe that $e^{(d)} \succ e^{(d-1)}\succ \cdots  \succ e^{(2)}\succ e^{(1)}$.
Now setting $\Sigma_k ( \rho )= \Sigma_k ( spec ( \rho ))$, $k=1,2,\cdots , d$,  these maps are convex continuous functions on ${\cal S}_d$.
One says that the state $\rho$ is more mixed (more chaotic) than the state $\phi$, and writes  $\rho \succ \phi $, if $\Sigma_k ( \rho ) \leq \Sigma_k ( \phi )$ for $k=1,2, \cdots , d$.
An intrinsic characterization is provided by Uhlmann's Theorem: $\rho \succ \phi$ iff $\rho \in co \{ \phi_u : \; u \in {\cal B}_d \mbox{ unitary}\}$. Another very useful characterization is:
\begin{equation} \mbox{$\rho \succ \phi$ iff $F( \rho ) \leq F( \phi )$} \;,\label{MONOTONIE}
\end{equation}
for every unitarily invariant, convex continuous $F$.\\

The non-increasing ordering of the components of an $\lambda \in {\cal L}_d$ imposes a number of bounds on the components of $\lambda$ and on the partial sums $\Sigma_k ( \lambda )$. These are particularly inmediate if one uses the (baricentric) coordinates $x_j$, $\lambda=\sum_{j=1}^d x_j e^{(j)}$, of $\lambda$. The following is an example
\begin{lema} \label{ABS} For $\lambda =(\lambda_1, \cdots , \lambda_d )\in {\cal L}_d$, one has:
\begin{enumerate}
\item $ d^{-1} \leq \lambda_1 \leq 1$; with equality in the left-hand side inequality iff $\lambda=e^{(d)}$, and in the right-hand side inequality iff $\lambda =e^{(1)}$. For each $k \in \{2,\cdots , d\}$, $0 \leq \lambda_k \leq k^{-1}$ with equality on the left-hand side inequality iff $\lambda\in co (e^{(1)}, \cdots , e^{(k-1)})$, and in the right-hand side inequality iff $\lambda =e^{(k)}$
\item For each $k \in \{1,2,\cdots , d\}$, $ kd^{-1} \leq \Sigma_k (\lambda ) \leq 1$. For $k < d$ one has equality on the left-hand side inequality iff $\lambda= e^{(d)}$, and in the right-hand side inequality iff $\lambda \in co ( e^{(1)}, \cdots , e^{(k)})$.
\end{enumerate}
\end{lema}

\dem  Let $\lambda = \sum_{j=1}^dx_j e^{(j)}$; one has $x_j \geq 0$ and $\sum_{j=1}^d x_j=1$. Since $\lambda_k =\sum_{j=k}^d x_j/j$ one has
\[  d^{-1}\sum_{j=k}^d x_j \leq \sum_{j=k}^d x_j/j \leq k^{-1}\sum_{j=k}^d x_j \leq k^{-1}\;.\]

One has $\Sigma_k ( \lambda ) = \sum_{j=1}^k x_j +k \sum_{j=k+1}^d x_j/j$ where the second sum is absent if $k=d$. Thus
\[  kd^{-1}= kd^{-1}(\sum_{j=1}^k x_j + \sum_{j=k+1}^d x_j ) \leq kd^{-1}\sum_{j=1}^kx_j+ k \sum_{j=k+1}^d x_j/j\]
\[ \leq \sum_{j=1}^kx_j+ k \sum_{j=k+1}^d x_j/j
 \leq \sum_{j=1}^kx_j+ k(k+1)^{-1} \sum_{j=k+1}^d x_j \]
 \[ = 1 - (k(k+1))^{-1} \sum_{j=k+1}^d x_j \leq 1 \;.\]
The inequality $\Sigma_k ( \lambda ) \leq 1$ is strict unless $\sum_{j=k+1}^d x_j =0$, that is to say $\sum_{j=1}^kx_j =1$, or $\lambda \in co ( e^{(1)}, \cdots , e^{(k)})$. The inequality $kd^{-1}\leq \Sigma_k ( \lambda )$ is strict unless
$k=d$, or $x_d =1$.$\Box$\\

\subsection{The gap-representation of a state}

Given a state $\rho$ with $spec ( \rho ) = (\lambda_1, \lambda_2, \cdots , \lambda_d) \in {\cal L}_d$ consider an orthonormal basis $\{ \psi_j : \; j=1,2, \cdots , d \}$ of ${\cal H}_d$ consisting of eigenvectors $\psi_j$ of the density matrix associated to $\rho$. Then
\[ \rho = \sum_{j=1}^d \lambda_j \cdot \rho^{(j)} \;,\]
where the pure states $\rho^{(j)}$ are given by $\rho^{(j)} (a) = \langle \psi_j, a \psi_j \rangle$, $a \in {\cal B}_d$.
This corresponds to a {\em  spectral decomposition} into pairwise orthogonal pure states which is not unique if $\rho$ has degenerate non-zero eigenvalues.

\begin{propo} \label{GR}Let $\{ \rho^{(j)}:\; j=1,2, \cdots , d\}$ be a maximal family of pairwise orthogonal pure states of ${\cal B}_d$. The set $\{ \sum_{j=1}^d \lambda_j \cdot \rho^{(j)} : \; (\lambda_1, \lambda_2 , \cdots , \lambda_d ) \in {\cal L}_d\}$ is affinely homeomorphic to ${\cal L}_d$. Thus it is a compact convex subset of the state space of ${\cal B}_d$ and a $(d-1)$-simplex with the $d$ extremal points given by the states
\[ \widehat{\rho}^{(j)} = j^{-1} \sum_{k=1}^j \rho^{(k)} \;,\; j=1,2, \cdots , d \;.\]
One has $\widehat{\rho}^{(d)}=\tau_d$, and
\begin{equation} \sum_{j=1}^d \lambda_j \cdot \rho^{(j)} = \sum_{j=1}^{d-1} \mu_j (\lambda)  \cdot \widehat{\rho}^{(j)} + d \lambda_d \cdot \tau_d \;,\label{gaprep} \end{equation}
where
\begin{equation}
\mu_j ( \lambda ) = j (\lambda_j -\lambda_{j+1}) \geq 0 \;,\;\; j=1,2, \cdots , d-1 \;,\label{gapcoeff}
\end{equation}
and $\sum_{j=1}^{d-1}\mu_j ( \lambda ) = 1-d\lambda_d$.
\end{propo}

\dem  If $0\leq t \leq 1$, and $\rho , \phi \in \{ \sum_{j=1}^d \lambda_j \rho^{(j)} : \; (\lambda_1, \lambda_2 , \cdots , \lambda_d ) \in {\cal L}_d\}$ with $spec ( \rho)=\lambda$ and $spec ( \phi )= \mu$, we have
\[ t\cdot \rho +(1-t)\cdot \phi = t\cdot \sum_{j=1}^d\lambda_j \cdot \rho^{(j)} + (1-t)\cdot \sum_{j=1}^d\mu_j \cdot \rho^{(j)}\]
\[ = \sum_{j=1}^d (t\lambda_j +(1-t)\mu_j)\cdot \rho^{(j)}\,\]
and thus $spec ( t \cdot \rho +(1-t)\cdot \phi )= t \cdot spec(\rho)+(1-t)\cdot spec ( \phi )$. The converse is also true, and $spec$ is an affine homeomorphism from  $ \{ \sum_{j=1}^d \lambda_j \cdot \rho^{(j)} : \; (\lambda_1, \lambda_2 , \cdots , \lambda_d ) \in {\cal L}_d\}$ onto  ${\cal L}_d$. Apply Proposition \ref{LD}.$\Box$\\

We call the representation of a state $\rho$ given by (\ref{gaprep}) the {\em gap-representation} of $\rho$ due to formula (\ref{gapcoeff}) which involves the successive eigenvalue gaps. This representation is not unique in as much as the spectral decomposition is not unique when one has spectral degeneracies. But any multiplicities if present, are automatically taken care of by the states $\widehat{\rho}^{(j)}$. Notice that $\widehat{\rho}^{(j)}\widehat{\rho}^{(k)}=
(1/\max \{j,k\})\widehat{\rho}^{(\min\{j,k\})}$. These algebraic equations are characteristic for a gap-representation as follows:
\begin{lema} If $k\in \{1,2, \cdots , d\}$ and $\alpha_1,\alpha_2, \cdots , \alpha_k$ are $k$ distinct states with
\begin{enumerate}
\item For every $j\in \{1,2,\cdots , k\}$ there is $s(j)\in \{1,2, \cdots , d\}$ such that $j_1\neq j_2$ implies $s(j_1)\neq s (j_2)$; i.e., $s$ is an injection from $\{1,2,\cdots , k\}$ into $ \{1,2, \cdots , d\}$;
\item $\alpha_j \alpha_m = (1/\max\{s(j),s(m)\})\alpha_{ s^{-1}(\min\{ j,m \})}$ for every $j,m \in \{1,2,\cdots , k\}$;
\end{enumerate}
then there is a maximal family $\{ \rho^{(n)}: \; n=1,2, \cdots , d\}$ of pairwise orthogonal pure states such that
$\alpha_j = \widehat{\rho}^{(s(j)}$ for every $j=1,2, \cdots , k$.
\end{lema}

\dem The $s(j)$'s are all distinct and the $s(j)\alpha_j=P_{s(j)}$ are orthoprojectors of rank $s(j)$ which satisfy $P_{s(j)}P_{s(m)}=P_{\min\{s(j),s(m) \}}$. Renumerate the $\alpha_j$'s so that  $1\leq s(1) < s(2) < \cdots < s(k)\leq d$; then
$P_{s(j)}P_{s(m)} = P_{s(\min\{j,m\})}$. Choose an orthonormal set $\{ \psi_n :\; n=1,2,\cdots, s (1)\}$ spanning the range of $P_{s(1)}$, an succesively orthonormal sets
 $\{\psi_n : \; n=s(j)+1, \cdots , s(j+1)\}$ spanning the range of $P_{s(j+1)}-P_{s(j)}$, for $j=1,2,\cdots , k-1$. Finally, choose an orthonormal set
 $\{ \psi_n: \; n=s(k)+1, \cdots, d\}$ spanning the kernel of $P_{s(k)}$. Then if $\rho^{(n)}$ denotes the pure state associated to the vector $\psi_n $, $n=1,2,\cdots , d$, we have $\alpha_j = \widehat{\rho}^{(s(j))}$.$\Box$\\

The gap-representation has a number of features which turn out to be useful in the discussion of entanglement.
The states $\widehat{\rho}^{(j)}$, which are the vertices of the $(d-1)$-simplex, obtained from different maximal families of pairwise orthogonal pure states are unitarily equivalent. For $j <d$ they have only two eigenvalues $0$ (with multiplicity $d-j$) and $1/j$ (with multiplicity $j$).
This will considerably simplify the discussion of their separability in composite systems.  \\

\subsection{Unitarily invariant convex functions on ${\cal S}$}

\label{UICF}

Consider a real-valued function $F$ defined on ${\cal S}_d$ which is unitarily invariant,  convex, i.e. $F(t\cdot \rho +(1-t)\cdot \phi ) \leq tF(\rho )+(1-t)F( \phi )$, for every $t\in [0,1]$ and every $\rho , \phi \in {\cal S}_d$, and continuous\footnote{In the present context lower semicontinuous convex functions are automatically continuous.}.
Let $F_+ := \sup \{ F( \rho ) : \, \rho\in {\cal S}_d\}$; by continuity and compactness  there is a maximizer.

\begin{propo} \label{F} If $F: {\cal S}_d \to {\mathbb R}$ is a unitarily invariant convex function then:
\begin{enumerate}
\item For every $\rho \in {\cal S}_d$,
\begin{equation}  F( \tau_d ) \leq F( \rho ) \leq F_+  \;.\label{range}
\end{equation}
Moreover $F (\rho )=F_+$ for every pure state $\rho$.
\item For each $c \in [F( \tau_d), F_+]$ the level set ${\mathbb L}_c:= \{ \rho \in {\cal S} : \; F( \rho ) \leq c \}$ is a compact, convex, unitarily invariant subset of ${\cal S}$.  If $\rho \succ \phi \in {\mathbb L}_c$ then $\rho \in {\mathbb L}_c$. Moreover,  ${\mathbb L}_c \subseteq {\mathbb L}_b$ if $c < b$.
\item If $F$ is strictly convex then there is equality in the left-hand side inequality of Eq. (\ref{range}) iff $\rho =\tau_d$; and there is equality in the right-hand side inequality of Eq. (\ref{range}) iff $\rho $ is pure. Moreover,
$ext ({\mathbb L}_c) = \{ \rho \in {\cal S} : \; F( \rho )=c \}$.
\end{enumerate}
\end{propo}

\dem The reader is asked to verify the triviality of (i) \& (ii) for a constant $F$. We thus assume that $F$ is not constant.\\

  Since, for every state $\rho$ and every pure $\phi$, one has  $\tau_d \succ \rho \succ \phi$ , the  inequality (\ref{range}) follows from Eq (\ref{MONOTONIE}), and $F_+ =F( \phi )$. Suppose $F$ is strictly convex and $\rho \neq \tau_d$ satisfies $F( \rho )=F( \tau_d)$, then $F( \tau_d)\leq F( t\cdot \rho +(1-t)\cdot \tau_d ) < tF(\rho )+ (1-t)F( \tau_d)=F( \tau_d)$ a contradiction. This  and the definition of $F_+$ proves Eq. (\ref{range}) and part of the statement of (iii). \\

(ii) is clear.\\

Suppose  $F$ is strictly convex.  If $\rho\in {\mathbb L}_c$ but  $\rho\notin ext ({\mathbb L}_c)$; then $\rho = t\cdot \phi+(1-t)\cdot \omega$ with $0<t<1$, $\phi , \omega \in {\mathbb L}_c$ and  $\phi\neq \omega$; thus  $F( \rho )< t F(\phi )+(1-t)F(\omega) \leq c$; this proves $\{ \rho : \; F( \rho )=c\}\subseteq ext ({\mathbb L}_c)$. For $c=F_+$ we conclude that $F(\rho )=F_+$ implies that $\rho$ is pure.  It remains to show that $ext ({\mathbb L}_c )\subseteq \{\rho : \; F( \rho )=c \} $. If $c=F( \tau_d)$ then ${\mathbb L}_{F( \tau_d)}= \{ \tau_d \}=\{\rho : \; F( \rho )=F( \tau_d )\}$. We assume that $c > F( \tau_d)$ and $F( \rho )<c$ and show that $\rho$ is not extremal in ${\mathbb L}_c$. Take any gap-representation of $\rho $, then $\rho =t_o\cdot \widehat{\rho}^{(1)} +(1-t_o)\cdot \omega $ with $spec ( \omega  )\in co (e^{(2)},\cdots , e^{(d)})$ and $t_o\in [0,1]$. The map $[0,1] \ni t \mapsto f(t)=F( t\cdot \widehat{\rho}^{(1)}+(1-t)\cdot \omega$ is strictly convex since, for $u, t_1,t_2\in [0,1]$
\[ f_{\phi } ( ut_1+(1-u)t_2) \]
\[ = F( (ut_1+(1-u)t_2)\cdot \widehat{\rho}^{(1)} +(1-(ut_1+(1-u)t_2))\cdot \omega )\]
\[ = F( u\cdot (t_1\cdot \widehat{\rho}^{(1)}+(1-t_1)\cdot \omega )+ (1-u)\cdot ( t_2\cdot \widehat{\rho}^{(1)}+(1-t_2)\cdot \omega)
\]
\[ \leq u F(t_1\cdot \widehat{\rho}^{(1)}+(1-t_1)\cdot \omega )+(1-u)F(t_2\cdot \widehat{\rho}^{(1)}+(1-t_2)\cdot \omega )
\]
\[ = u f(t_1)+(1-u)f(t_2)\;,\]
and the inequality is strict if $0<u<1$ and $t_1\neq t_2$ by the strict convexity of $F$ and the fact that $t_1\cdot \widehat{\rho}^{(1)}+(1-t_1)\cdot \omega\neq t_2\cdot \widehat{\rho}^{(1)}+(1-t_2)\cdot \omega$. Moreover, if $1\geq t_1 > t_2\geq 0$ then
$t_2\cdot \widehat{\rho}^{(1)} +(1-t_2)\cdot \omega \succ t_1\cdot\widehat{\rho}^{(1)} +(1-t_1)\cdot \omega$ since this is equivalent to
 $(t_1-t_2)(\Sigma_k ( \widehat{\rho}^{(1)}) -\Sigma_k ( \omega ))\geq 0$, and the latter follows from $\omega \succ \widehat{\rho}^{(1)}$. But then by Eq. (\ref{MONOTONIE}), $f(t_2)=F(t_2\cdot \widehat{\rho}^{(1)} +(1-t_2)\cdot \omega )\leq F(t_1\cdot \widehat{\rho}^{(1)} +(1-t_1)\cdot \omega )=f (t_1)$ so that $f_{\phi}$ is non-decreasing. But a strictly convex, non-decreasing function must be increasing. Thus $F( \omega )=f(0)\leq f(t_o)=F(\rho )< c \leq F_+$; let $t_*$ be the unique number in $[0,1]$ such that $f(t_*)=c$ it follows that $t_o < t_*$ and thus  $\rho = t_o \cdot \widehat{\rho}^{(1)}+(1-t_o )\cdot \omega = (t_o/t_*)\cdot (t_*\cdot \widehat{\rho}^{(1)} +(1-t_*)\cdot \omega )+( 1-(t_o/t_*))\cdot \omega  $ is not extremal in ${\mathbb L}_c$ since $f(0)=F( \omega )<c$ and $F(t_*\cdot \widehat{\rho}^{(1)}+(1-t_*)\cdot \omega )=f(t_*)=c$.$\Box$\\

We will not make much use of what follows but record it for completeness.\\

Suppose $F: {\cal S}_d \to {\mathbb R}$ is unitarily invariant. Then $f_F ( \lambda )= F ( \rho )$, $spec ( \rho )=\lambda \in {\cal L}_d$ is well defined and is continuous if $F$ is.  Suppose $F$ is also convex. For $\lambda , \mu \in {\cal L}_d$ and $t\in [0,1]$, take any maximal family $\{ \rho^{(j)}:\; j=1,2, \cdots , d\}$ of pairwise orthogonal pure states, and put $\phi = \sum_{j=1}^d \lambda_j \cdot \rho^{(j)}$, $\omega = \sum_{j=1}^d \mu_j\cdot \rho^{(j)}$, and
$ \rho = t\cdot \phi +(1-t)\cdot \omega= \sum_{j=1}^d (t\lambda_j +(1-t)\mu_j)\cdot \rho^{(j)}$. Then, $spec( \phi)=\lambda$, $spec ( \omega )=\mu$ and $spec( \rho )=t\cdot \lambda +(1-t)\cdot \mu$ because the nonincreasing ordering of $spec$. Thus,
\[ f_F ( t\cdot \lambda +(1-t)\cdot \mu )= F( \rho )= F( t\cdot \phi +(1-t)\cdot \omega )\]
\[ \leq tF( \phi )+(1-t)F( \omega) =
t f_F( \lambda )+(1-t)f_F(\mu )\;,\]
and the inequality is strict if $0<t<1$, $\lambda \neq \mu$ and $F$ is strictly convex. This shows that $F\to f_F$ transforms unitarily invariant, convex and continuous functions on ${\cal S}_d$ into convex continuous functions on ${\cal L}_d$ such that strictly convex $F$'s give strictly convex $f_F$'s.\\

Conversely, if $f$ is any real-valued function on ${\cal L}_d$, then $F_f( \rho )=f( spec ( \rho ))$, $\rho \in {\cal S}_d$, gives a real-valued unitarily invariant function on ${\cal S}_d$ which is continuous if $f$ is.
Let $f( \lambda )= \lambda_d$ then, due to the non-increasing enumeration of the components of vectors in ${\cal L}_d$,  $f(t \cdot \lambda +(1-t)\cdot \mu )=t \lambda_d +(1-t)\mu_d$; so $f$ is affine. However, $F_f$ is not convex. Indeed,
$\tau_d = \sum_{j=1}^d (1/d) \rho^{(j)}$ for any maximal family $\{ \rho^{(j)}:\; j=1,2, \cdots , d\}$ of pairwise orthogonal pure states; but $F_f ( \rho^{(j)} )= f( spec ( \rho^{(j)}))=f( e^{(1)})=0$ so that $0= \sum_{j=1}^d (1/d)F_f ( \rho^{(j)}) < (1/d)=F_f ( \tau_d)=
F_f ( \sum_{j=1}^d (1/d) \rho^{(j)} )$.

\section{Spectral conditions implying separability}

We return to the discussion of arbitrary compositions of finite quantum systems as described in the Introduction. Given integers $d_1, d_2, \cdots , d_N$ all of which are larger or equal to $2$, and $N$ Hilbert spaces ${\cal H}_j$ of dimension $d_j$, consider ${\cal H}= {\cal H}_1\otimes {\cal H}_2 \otimes \cdots \otimes {\cal H}_N$ which has dimension $D=d_1d_2\cdots d_N$.   We identify  $ {\cal B} = {\cal B}({\cal H}) $ with  ${\cal B}_{d_1}\otimes {\cal B}_{d_2} \otimes \cdots \otimes {\cal B}_{d_N}$.
A state $\rho$ of ${\cal B} $ is {\em separable} if it lies in the convex hull of the product states of ${\cal B} $; otherwise it is called {\em entangled}. ${\cal S}^{sep}$ denotes the separable states.\\

 The following result is an inmediate consequence of Uhlmann's Theorem, and proves to be  useful if one has a rich zoo of states $\rho$ for which $\rho_u$ is separable for all unitary $u \in {\cal B} $.

\begin{propo} \label{hered} If $\rho \in {\cal S}$ is such that $\rho_u $ is separable for every unitary $u \in {\cal B} $ and $\phi \in {\cal S}$ satisfies $spec ( \phi ) \succ spec ( \rho)$ then $\phi$ is separable.
\end{propo}

We begin to populate the zoo in the following subsection.

\subsection{The separability-modulus. Generalities}
We  observe that $\tau_D=\tau_{d_1} \otimes \tau_{d_2} \otimes \cdots \otimes \tau_{d_N}$ so that $\tau \equiv \tau_D$ is a product-state hence separable. $\tau $ is also the maximally mixed state of ${\cal B} $.
Given any state $\rho $ of ${\cal B} $ and any $t \in [0,1]$, we let
\[ \rho_{t} = (1-t ) \cdot \tau + t \cdot  \rho \;.\]
We then ask ourselves: ?`when is $\rho_t$ separable?

Frequently in what follows we use the fact that if  $\omega , \varphi$ are both separable states then $t\cdot\omega +(1-t)\cdot\varphi$ is separable for every  $t \in [0,1]$ because ${\cal S}^{sep}$ is convex. \\

We observe that if  $\rho$ is separable then $\rho_{t}$ is separable for every  $t \in [0,1]$. In fact,
\begin{lema} $\,$
\begin{enumerate}
\item $\rho_{t}$ is separable for every  $t \in [0,1]$ iff  $\rho$ is separable;
\item if $\rho_{t}$ is separable for some  $t \in (0,1]$ then  $\rho_s$ is separable for every $s \in [0, t]$;
\item if  $\rho_{s}$ is entangled  for some  $s \in (0,1]$ then  $\rho_{t}$ is entangled for every $t \in [s , 1]$.
\end{enumerate}
\end{lema}

\noindent \underline{Proof}: 1. is clear. Suppose $0\leq s \leq t\leq 1$, then $0\leq s/t\leq 1$ and
\[ \rho_s = s \cdot \rho +(1- s )\cdot \tau = \frac{s}{t} \cdot \left(t \cdot \rho +(1- t )\cdot \tau \right) + \left( 1 - \frac{s}{t}\right) \cdot\tau\]
\[ = \frac{s}{t}\cdot \rho_{t} + \left( 1 - \frac{s}{t}\right) \cdot \tau\;.\]
Under the hypotheses of 2., $\rho_s$ is separable as a convex sum of two separable states. Under the hypotheses of 3., $\rho_s$ would be separable if $\rho_t$ were separable.$\Box$\\

The above allows us to introduce the {\em modulus of separability} of $\rho$ (with respect to $\tau$) as the number
\[ \ell (\rho) = \sup \{ t \in [0,1]:\; \rho_{t} \mbox{ is separable }\}\;.\]

Vidal and Tarrach \cite{ViTa} have studied the quantity  $\ell(\rho) ^{-1}-1$ which they called the {\em random robustness of entanglement}. Most of the results below are explicitely or implicitely given by them so the rest of this section is a streamlined exposition of the basic facts about $\ell$ that we need. For more information the reader should consult \cite{ViTa}.\\

\begin{lema} If  $t \in [0,1]$ then $\ell ( \rho_t ) = \min \{ 1 , t^{-1} \ell ( \rho )\}$ for every state $\rho$. \label{l_t}
\end{lema}

\noindent \underline{Proof}:  A straightforward calculation  gives $(\rho_t)_s = \rho_{ts}$. Suppose that $t >0$; then
\[ \ell ( \rho_t ) = \sup \{ s \in [0,1] : \;  (\rho_t)_s \mbox{  es separable}\} \]
\[ =  \sup \{ s\in [0,1] :\; \rho_{ts} \mbox{ es separable}\} \]
\[ = \sup \{ r /t \in [0,1] : \; \rho_r \mbox{  es separable}\}\]
\[ = t^{-1} \sup \{ r \in [0, t] : \rho_r  \mbox{  es separable}\}\;.\]
If $t < \ell ( \rho )$ the last supremum is  $t$. If $ \ell (\rho ) \leq t$ the last supremum is  $\ell ( \rho )$.
With the usual interpretation, the formula remains valid for $t=0$ since $\rho_0=\tau$ and  $\ell ( \tau )=1$.$\Box$\\

\begin{lema} $\rho_{\ell ( \rho )}$  is separable. \label{l_l}
\end{lema}

\noindent \underline{Proof}: For $t <\ell ( \rho )$ we have  $\rho_{\ell ( \rho )}-\rho_t = (\ell (\rho ) -t) \cdot (\rho - \tau )$ and thus  $\parallel\rho_{\ell ( \rho )}-\rho_t \parallel = (\ell (\rho ) -t) \parallel \rho - \tau \parallel$. Taking a sequence $\{ t_n : \; n=1,2, \cdots \}$ with $t_n < \ell ( \rho )$ and  $\lim_{n \to \infty} t_n =\ell ( \rho )$, we have $\rho_{t_n} \in {\cal S} ^{sep}$ and $\lim_{n \to \infty} \rho_{t_n} = \rho_{\ell (\rho ) }$ and hence $\rho_{\ell (\rho )}$ lies in the closure of ${\cal S} ^{sep}$ which is closed.$\Box$\\

\begin{cor} $\rho_t$ is separable iff $t \leq \ell ( \rho )$.
\end{cor}

 \begin{lema} If $0 < t_j \leq 1$ for $j=1,2,\cdots , M$ with $M \in {\mathbb N}$, and  $\sum_{j=1}^M t_j =1$, then
\[  \ell \left( \sum_{j=1}^M t_j \cdot \rho^{(j)} \right) \geq \min \{ \ell (  \rho^{(j)} ): \; j=1,2, \cdots , M \} \]
for any set of  $M$ states $\{  \rho^{(j)} :\; j=1,2,\cdots ,M\}$. \label{min}
\end{lema}

\noindent \underline{Proof}: The following proof does not use the previous result. Let $\omega = \sum_{j=1}^M t_j \cdot  \rho^{(j)} $; then
\[ \omega_{t} = t \cdot \omega +(1-t )\cdot \tau =
\sum_{j=1}^M t_j (\rho^{(j)})_{t}\;.\]
If  $t < t_o:= \min \{ \ell (  \rho^{(j)} ): \; j=1,2, \cdots , M\}$, the states  $(\rho^{(j)})_{t}$ are all separable  and thus
$\omega_{t}$ is separable; by definition of  $\ell$, $\ell ( \omega ) \geq t$. Taking the supremum with respect to  $t < t_o$ one obtains the result.$\Box$\\

A substantial improvement of the above lower bound would be concavity of $\ell$. Examples for two qubits show that this is not the case. However, $ \rho \mapsto (1/\ell ( \rho ))$ turns out to be convex.\\

 Consider
 \[ L := \inf \{ \ell (\rho ) :\; \rho \in {\cal S}\}\;;\]
 due to Lemma \ref{min}, the infimum can be taken over the pure states. Moreover, $L <1$ since there are entangled pure states. $L$ has been computed in various cases \cite{ViTa,Ru,Bra}.
 For our purposes it would  suffice to know that $L >0$, and \.{Z}yczkowski et al., \cite {Zy}, give an elegant proof of this.    Rungta, \cite{Ru}, for $d_1=d_2=\cdots =d_N$ (using the methods of \cite{Bra}) obtained\footnote{To handle the case of distinct dimensions $d_j$ we just embed ${\cal B}_{d_j}$ in ${\cal B}_d$  by adding the necessary rows and columns of zeroes.}:
\[ L \geq \frac{1}{1+d^{2N-1}}\;,\;\; d=\max \{ d_1,d_2, \cdots , d_N\}\;;
\]
while Vidal and Tarrach, \cite{ViTa}, obtained:
\[ L \geq \frac{1}{(1 + D/2)^{N-1}}\;.\]

The second bound is exact for $N=2$ while the Rungta bound is poor for this case. As $N$ increases, the first bound eventually exceeds the second one and becomes the better lower bound on $L$.

\begin{propo} $1/\ell$ is convex. \label{conv}
\end{propo}

\noindent \underline{Proof}: Convexity of $1/\ell$ is
\[  \ell ( t\cdot\rho^{(1)}+(1-t)\cdot\rho^{(2)})^{-1} \leq t\ell ( \rho^{(1)})^{-1} +(1-t)\ell ( \rho^{(2)})^{-1} \;,\]
or equivalently
\[   \ell (t \cdot \rho^{(1)} +(1-t)\cdot\rho^{(2)}) \geq  \frac{\ell ( \rho^{(1)}) \ell ( \rho^{(2)})}{t\ell ( \rho^{(2)})+(1-t)\ell ( \rho^{(1)})}\;.\]
Put $\ell_j:=\ell ( \rho^{(j)})$ and $s=\ell_1\ell_2/(t\ell_2+(1-t)\ell_1)$. Observe that $s \geq 0$ and $s\leq \ell_1\ell_2/\max\{\ell_1 ,\ell_2\}=\min\{\ell_1 , \ell_2\}\leq 1$. Thus, by Corollary 1, convexity is proved if we show that $(t\cdot \rho^{(1)}+(1-t)\cdot\rho^{(2)})_s$ is separable. But
\[ (t\cdot\rho^{(1)}+(1-t)\cdot\rho^{(2)})_s = s\cdot (t\cdot\rho^{(1)}+(1-t)\cdot\rho^{(2)})+(1-s)\cdot\tau \]
\[ = \left(\frac{\ell_1\ell_2t}{t\ell_2+(1-t)\ell_1}\right)\cdot  \rho^{(1)}+\left(\frac{\ell_1\ell_2(1-t)}{t\ell_2+(1-t)\ell_1}\right)\cdot \rho^{(2)}+(1-s)\cdot\tau\;.\]
Put
\[ r:= \left( \frac{\ell_2t}{t\ell_2+(1-t)\ell_1}\right) \;;\]
which is in $[0,1]$. Then
\[ 1-r = \left( \frac{\ell_1(1-t)}{t\ell_2+(1-t)\ell_1}\right) \;,\;\; (1-s)=1-r\ell_1-(1-r)\ell_2\;,  \]
so that
\[ (t\cdot \rho^{(1)}+(1-t)\cdot \rho^{(2)})_s = r \ell_1 \cdot\rho^{(1)} \]
\[ + (1-r)\ell_2\cdot \rho^{(2)} + (1-r\ell_1-(1-r)\ell_2).\tau\]
\[ =
r\cdot ( \ell_1\cdot \rho^{(1)}+(1-\ell_1 )\cdot \tau) +(1-r)\cdot (\ell_2\cdot\rho^{(2)}+(1-\ell_2 )\cdot\tau ) \]
\[ = r\cdot (\rho^{(1)})_{\ell_1} +(1-r)\cdot (\rho^{(2)})_{\ell_2}\;,\]
which is separable by Lemma \ref{l_l}.$\Box$\\

The convexity of $1/\ell$ gives an improvement on the lower bound of Lemma \ref{min}:

\begin{cor} If $0 < t_j \leq 1$ for $j=1,2,\cdots , M$ with $M \in {\mathbb N}$, and  $\sum_{j=1}^Mt_j =1$, then
\[  \ell \left( \sum_{j=1}^M t_j \cdot \rho^{(j)} \right) \geq \left( \sum_{j=1}^M \frac{t_j}{\ell (\rho^{(j)})} \right)^{-1} \geq    \min \{ \ell (  \rho^{(j)} ): \; j=1,2, \cdots , M \} \]
for any set of  $N$ states $\{  \rho^{(j)} :\; j=1,2,\cdots ,M\}$. \label{lower}
There is equality in the right-hand-side inequality iff all $\ell ( \rho^{(j)})$'s are equal.
\end{cor}

Although the convexity bound is saturated when all $\rho^{(j)}$'s are separable it is a rather poor bound when at least one of the $\rho^{(j)}$'s is entangled as we will see below.\\

\begin{propo} $\ell$ is upper semicontinuous.
\end{propo}

\noindent \underline{Proof}: We have to show that the sets ${\cal K}_x=\{ \rho :\; \ell (\rho ) \geq x\}$ are closed for every real $x$. These sets are empty for $x >1$ and are the whole space of states for $x\leq 0$. Otherwise, for $x \in (0,1]$ we have $\rho_x \in {\cal S} ^{sep}$ for every $\rho \in  {\cal K}_x$. If the sequence $\{\rho^{(n)}:\; n=1,2, \cdots \} \subset {\cal K}_x$ converges to $\rho$ then the sequence $\{ \rho^{(n)}_x : \; n=1,2, \cdots \}$ is in ${\cal S} ^{sep}$ and converges to $\rho_x$. Thus, since ${\cal S} ^{sep}$ is closed, $\rho_x$ is separable and thus $\ell ( \rho )\geq x$ so $\rho \in {\cal K}_x$.$\Box$\\

Thus $\rho \mapsto E( \rho):= (1/\ell ( \rho ))-1$ is a bonafide measure of entanglement in as much as  it is convex and lower semicontinuos, and it is zero iff $\rho$ is separable. As mentioned, Vidal and Tarrach \cite{ViTa} have studied $E$ extensively, and computed it in a number of particular cases.\\

Uppersemicontinuity of $\ell$ and compactness of ${\cal S}$ imply that there exists a state $\rho$ such that $\ell ( \rho )=L$. Any state with this property will be called {\em maximally entangled} and is automatically pure\footnote{The relationship with other notions of what a maximally entangled pure state is, need yet to be explored. The pure state given by the vector $(1/\sqrt{d})\sum_{j=1}^d \psi_j\otimes \psi_j$ for $N=2$ and $d_1=d_2=d$, does not seem to be maximally entangled in the above sense for $d>2$.}.\\

\subsection{Putting the gap-representation to work}

We can now use the gap-representation to obtain our weakest result of the type described by the title of the paper:

\begin{teor}  If $s_- ( \rho ) \geq (1-L)/D$ then $\rho$ is separable. For every $s \in [0,(1-L)/D)$ there is an entangled state $\phi$ such that $s_- ( \phi)=s$.\label{ULALA}
\end{teor}

\dem  Let $spec ( \rho ) = (\lambda_1, \lambda_2, \cdots , \lambda_D )$; 
if the condition on $\lambda_D =s_- ( \rho )$ is met  the gap-re\-pre\-sen\-ta\-tion gives
\[ \rho = \sum_{j=1}^{D-1} \mu_j ( \lambda ) \cdot \widehat{\rho}^{(j)} + D s_- ( \rho ) \cdot \tau
\]
\[ = (1- D s_- ( \rho ))\cdot \underbrace{\left(  \sum_{j=1}^{D-1} \frac{ \mu_j ( \lambda)}{ 1- D s_ ( \rho )} \,\cdot  \widehat{\rho}^{(j)}\right)}_{\phi } + Ds_- ( \rho )\cdot \tau \]
\[ = (1- D s_- ( \rho ))\cdot \phi + Ds_-(\rho ) \cdot \tau = \phi_{1-Ds_-(\rho )}\;;\]
but $1-Ds_-(\rho )\leq L$ and thus $(1-Ds_- (\rho)) \leq \ell ( \phi)$ which implies that $\rho=\phi_{1-Ds_-(\rho )}$  is separable.\\

Take any maximally entangled state $\omega$, then $\omega_t $ is separable iff $t \leq L$, and, because $\omega$ is pure,  $s_-( \omega_t)=(1-t)/D$. For any $s\in [0, (1-L)/D)$ the state $\omega_{1-Ds}$ is entangled and $s_( \omega_{1-Ds})=s$. $\Box$\\

We observe that, in the language of \S 2.1, the theorem states that \[ \{\rho :\; \Sigma_{D-1} ( \rho ) \leq (D+L-1)/D\}\subseteq {\cal S}^{sep}\;.\\ \]

An inmediate improvement of the weak result above would follow if it were true that: {\em every state $\rho$ with $spec ( \rho )= e^{(D-1)}$  is separable}. In Appendix A, we show that for $N=2$, $spec ( \rho )=e^{(D-1)}$ implies that $\rho$ has positive partial transpose, a condition which is known to be necessary for separability (\cite{HHH}).
How the separability of $\rho$'s with $spec ( \rho )=e^{(D-1)}$  can be used to improve Theorem \ref{ULALA} is shown in Appendix B.\\

Another useful feature of the gap-representation is that it provides an improvement on the convexity bound of Corollary \ref{lower} obtained from  any spectral decomposition of a state:

\begin{lema} Let $spec( \rho )= ( \lambda_1, \lambda_2 , \cdots , \lambda_D)\in {\cal L}_D$ and
\[ \rho = \sum_{j=1}^D \lambda_j \cdot \rho^{(j)} = \sum_{j=1}^{D-1} \mu_j ( \lambda ) \cdot  \widehat{\rho}^{(j)} + D\lambda_D . \tau \]
be a gap-representation. Then
\[ \ell (\rho ) \geq \left( \sum_{j=1}^{D-1} \frac{j(\lambda_j - \lambda_{j+1})}{\ell ( \widehat{ \rho}^{(j)})} +D \lambda_D \right)^{-1} \geq \left( \sum_{j=1}^D \frac{\lambda_j }{\ell ( \rho^{(j)})} \right)^{-1} \;.\]
\end{lema}

\dem Let us abreviate $\ell ( \rho^{(j)} ) = \ell_j$ and $\ell ( \widehat{\rho}^{(j)} )= \widehat{\ell}_j$. Applying Corollary \ref{lower} to the gap-representation and observing that $\widehat{\ell}_D=1$ we get
\[ \ell ( \rho ) \geq \left( \sum_{j=1}^{D-1} \frac{\mu_j ( \lambda)}{\widehat{\ell}_j} + D\lambda_D \right)^{-1}\;.\]
The other inequality is equivalent to
\[ \sum_{j=1}^D \frac{\lambda_j }{\ell_j} \geq \sum_{j=1}^{D-1}\frac{\mu_j ( \lambda)}{\widehat{\ell}_j} + D\lambda_D\;,\]
which we now prove.  Another application of Corollary \ref{lower} gives
\[ 1/\widehat{\ell}_j \leq  j^{-1} \sum_{k=1}^j (1/ \ell_j) \;,\;\; j=1,2, \cdots , D \;;\]
thus
\[ \sum_{j=1}^{D-1}\frac{\mu_j ( \lambda)}{\widehat{\ell}_j} + D\lambda_D \leq \sum_{j=1}^{D-1} \frac{\mu_j ( \lambda)}{j} \left( \sum_{k=1}^j (1/\ell_k)\right)  + \lambda_D \left( \sum_{k=1}^D (1/\ell_k)\right)
\]
\[ = \sum_{j=1}^{D-1} (\lambda_j - \lambda_{j+1}) \left( \sum_{k=1}^j (1/\ell_j)\right)  + \lambda_D \left( \sum_{k=1}^D (1/\ell_k) \right) = \sum_{j=1}^D \lambda_j /\ell_j \;.\; \Box \\ \]

\subsection{An application to thermal states}

Given any selfadjoint $h \in {\cal B} $  and  $\beta \in {\mathbb R}$ consider the (thermal equilibrium) state $\rho_{\beta}$ given by  the density operator
\[ \rho_{\beta } = \left\{ \begin{array}{lcl}
\tau &,& \mbox{ if $\beta =0$}\\
\exp ( - \beta h )/tr( \exp (- \beta h)) &,& \mbox{ if $\beta \neq 0$}
\end{array} \right. \;\;.\]
One has $\lim_{\beta \to 0} \rho_{\beta} = \tau$ and the expectation is that for $|\beta |$ sufficiently small one will have separability of $\rho_{\beta}$. This is indeed the case.

\begin{propo} There are real numbers $\beta_c^{\pm}$ with $\beta_c^¯ < 0 < \beta_c^+$ such that $\rho_{\beta}$ is separable for $\beta \in [\beta_c^¯, \beta_c^+]$, and if $I$ is any interval that contains $[\beta_c^-, \beta_c^+]$ properly, then there is $\beta'\in I$ such that $\rho_{\beta '}$ is entangled.
\end{propo}

\dem Assume $\beta \neq 0$ and put $Z(\beta )= tr ( e^{-\beta h})= \sum_{j=1}^D e^{-\beta \epsilon_j}$, where $spec(h) = (\epsilon_1, \cdots , \epsilon_D)$. Take $\beta > 0$; then $Z(\beta ) \leq D e^{-\beta s_- (h)}$ where $s_- (h)$ is the minimal eigenvalue of $h$.  Now $ s_- ( \rho_{\beta} ) = e^{-\beta s_+ (h)}Z(\beta )^{-1} \geq D^{-1}\exp ( -\beta (s_+ (h)-s_-(h))$. If
\[ \exp \{ -\beta (s_+ (h)-s_-(h))\}  \geq (1-L)\;,\]
the hypothesis of Theorem \ref{ULALA} is met and we conclude that $\rho_{\beta}$ is separable. But $s_+(h)>s_- (h)$ unless $h$ is a multiple of ${\bf 1}$ in which case $\rho_{\beta}=\tau$ for every $\beta$. Thus, $\rho_{\beta}$ will be separable if
\[ \beta \leq \beta_o := \frac{ -\ln \left( 1-L\right)}{s_+(h)-s_-(h)}\;.\]
A analogous argument in the case $\beta <0$ shows that $\rho_{\beta}$ is separable if
$\beta \geq -\beta_o$.  Define
\[ \beta_c^+ = \sup \{ \beta >0 : \; \rho_{\gamma} \mbox{ is separable for every $\gamma \in [0,\beta ]$}\}\; ,\]
\[ \beta_c^- = \inf \{ \beta <0 : \; \rho_{\gamma} \mbox{ is separable for every $\gamma \in [\beta,0 ]$}\} \;;\]
then $ \beta_c^- < -\beta_o <0 < \beta_o < \beta_c^+$. Moreover, since $\beta \to \rho_{\beta}$ is  continuous $\rho_{\beta_c^{\pm}} $ is separable as a  limit of separable states.$\Box$\\

Observe that $\beta_c^+=\infty$ and $\beta_c^-=-\infty$ are possible, e.g., when all spectral orthoprojectors of $h$ are products.\\

A beautiful result of Uhlmann and Wehrl (\cite{AlUh,OhPe}) says that $\beta \to \rho_{\beta}$ is $\succ$-decreasing for positive $\beta$ ($\succ$-increasing for negative $\beta$): $\rho_{\beta_1} \succ \rho_{\beta_2}$ if $0\leq \beta_1 < \beta_2$. The limit $\beta \to \pm \infty$ of $\rho_{\beta}$ is the state $\rho_{\pm \infty} = P_{\mp}/m_{\mp}$ where $P_-$ (resp. $P_+$) is the orthoprojection onto the eigenspace of the minimal eigenvalue $s_-$ (resp. the maximal eigenvalue $s_+(h)$) of the Hamiltonian and $m_-$ (resp. $m_+$) is its multiplicity. If $\rho_{\infty}$ is entangled (e.g., $m_-=1$ and the ground-state vector is not a product-vector), then $\beta_c^+ < \infty$.
What happens above $\beta_c$? This is under investigation and I will not venture a conjecture at present.\\

An observation of G. Toth, \cite{To}, can be used to give an upper (resp. lower) bound on $\beta_c^+$ (resp. $\beta_c^-$):
\begin{propo}  Let
\[ \eta_-= \inf \{ \rho (h) : \; \rho \in {\cal S}^{sep}\}\;;\;\; \eta_+ = \sup \{ \rho (h) : \; \rho \in {\cal S}^{sep}\}\;.\]
If the state $\rho_{\infty} = P_- /m_-$ is entangled then $\beta_c^+ \leq \beta_-$ where $\beta_-$ is the unique number in $(0, \infty )$ such that $\rho_{\beta_-}(h)=\eta_-$.
If the state $\rho_{-\infty}=P_+/m_+$ is entangled then $\beta_c^- \geq \beta_+$ where $\beta_+$ is the unique number in $(-\infty , 0)$ such that $\rho_{\beta_+}(h)= \eta_+$.
\end{propo}

\dem If $h$ is not a multiple of the identity, the map ${\mathbb R}\ni\beta \to U( \beta )= \rho_{\beta} (h)$ has derivative $U'( \beta )= - \rho_{\beta} ([h-\rho_{\beta}(h)]^2)$ and is thus decreasing. Consider the case of non-negative $\beta$; the other is analogous. We have $U(0)=\tau ( h) > \eta_-$, and if $\rho_{\infty}$ is entangled then $\lim_{\beta \to \infty} U( \beta ) =s_- < \eta_-$. The intermediate-value theorem gives a unique $\beta_-$ such that $U( \beta_- )= \eta$, and if $\beta > \beta_-$ then $\rho_{\beta}(h) < \eta_- $ and the definition of $\eta_-$ implies that $\rho_{\beta}$ is entangled.$\Box$\\
 
Notice that in the variational problem defining $\eta_{\pm}$, one can restrict oneself to pure product-states.\\

\subsection{Unitarily invariant convex functions as separability detectors} \label{D}

We use the definitions and notation of  \ref{UICF}.\\
A  function $F$ on the state space of a composite system is said to be {\em good} if $F(\rho )=F( \tau )$ implies that $\rho=\tau$. In particular, $F$ cannot be constant. By Proposition \ref{F}, every unitarily invariant strictly convex continuous function is good. Also, any $k$-th eigenvalue partial-sum $\Sigma_k$ with $k<D$ is good by Lemma \ref{ABS}.  \\

If $F: {\cal S} \to {\mathbb R}$ is unitarily invariant, convex and continuous, the numbers
\[  C_L^+ = F ( L\cdot \phi +(1-L)\cdot \tau ) \;,\;\; \mbox{ $\phi$ pure}\;;\]
\[ C_L^- =  F( L\cdot \omega +(1-L)\cdot \tau )\;,\;\;  spec (\omega )= e^{(D-1)}  \;,\]
are well defined. To see that $C_L^- \leq C_L^+$, take  an $\omega $ with $spec ( \omega )=e^{(D-1)}$ and  take any gap-representation of $\omega$; the state $\widehat{\rho}^{(1)}$ is pure and $\omega \succ\widehat{\rho}^{(1)} $ so that $C_L^-=F(L\cdot \omega +(1-L)\cdot \tau )\leq F(L\cdot \rho^{(1)}+ (1-L)\cdot \tau)=C_L^+$.    
 Also $C_L^- = F( L\cdot \omega+(1-L)\cdot \tau ) \geq F( \tau )$ because of Proposition \ref{F}, with strict inequality if $F$ is good; and $C_L^+ \leq L F( \phi )+ (1-L)F( \tau ) = LF_+ +(1-L) F( \tau ) \leq F_+$, with strict inequality for non-constant $F$. The reason for introducing these numbers is:\\

\begin{teor} \label{CRIT} If $F: {\cal S} \to {\mathbb R}$ be  a unitarily invariant, convex, continuous and good function, then  there is a number  $C_F$ in $[ C_L^-, C_L^+]$ such that  ${\mathbb L}_c \subset {\cal S}^{sep}$ for every $c\in [F(\tau),C_F]$, whereas for every
$c' \in (C_F, F_+]$ there is an entangled state $\rho$ with $F(\rho )=c'$.
\end{teor}

There is of course a version of the above for unitarily invariant,  concave,   continuous and good functions (i.e., entropies that deserve their name); the statement is then $\{ \rho :\; F ( \rho )\geq c\} \subseteq {\cal S}^{sep}$ iff $c\geq C_F$. The above result shows that every unitarily invariant continuous and good  function on ${\cal S}$ which is convex or concave provides us with a separability detector. The list of separabilty detectors includes, apart from the $k$-th partial-sums for $k=1,2,\cdots , D-1$,
the von Neumann entropy $S( \rho ) = tr (\rho \ln ( \rho))$, the R\'enyi entropies, the functions $F( \rho ) = tr ( \rho^q)$ with $q>0$, etc.   \\

The rest of this section is devoted to the proof of this Theorem, and to the computation of the bounds $C_L^{\pm}$ for some especially interesting $F$'s.\\

By Proposition \ref{F},  the level sets ${\mathbb L}_c$ are closed and convex for every $c\in [F(\tau ), F_+]$ and not empty since $\tau \in {\mathbb L}_c$ always. 
The basic observation is
\begin{lema} \label{>} If $F$ is good and $\tau \neq \rho \in {\cal S}$,  then the map $[0,1] \ni t \mapsto f_{\rho} (t) = F( t.\rho+(1-t).\tau)$  is increasing, and  convex.
\end{lema}

\dem $f_{\rho}$ is constant for $F (\rho )=F( \tau)$; otherwise, by convexity of $F$, it is convex and, by goodness, it  assumes its minimal value $F(\tau )$ precisely at $t=0$. It follows that the map is  increasing.$\Box$\\

Let $\rho$ be a maximally entangled state; then it is pure and $F(L.\rho +(1-L).\tau)=C_L^+$. By the above Lemma, for every $F_+\geq c > C_L^+$ there is a (unique) $t>L$ such that $F(t.\rho +(1-t).\tau)=f_{\rho}(t)=c$ and $t.\rho +(1-t).\tau$ is entangled. This shows that for every $c > C_L^+$, ${\mathbb L}_c$ is not contained in ${\cal S}^{sep}$ but contains an entangled state $\phi$ with $F (\phi )=c$. \\

Let $C=\{ c \in [F(\tau ), F_+]:\; {\mathbb L}_c \subseteq {\cal S}^{sep}\}$. Then $C$ is not empty because ${\mathbb L}_{F( \tau)}=\{ \tau \} \subseteq {\cal S}^{sep}$. Put $C_F = \sup (C)$ ($ \leq C_L^+$), and  ${\mathbb K}=\cup_{c \in C} {\mathbb L}_c$.\\

 We first show that $C_F \geq C_L^-$.
Suppose that $\rho \in {\mathbb L}_{C_L^-}$ then $\rho = t \cdot \phi + (1-t)\cdot \tau =\phi_t$ with $t \in [0,1]$ and $spec (\phi )\in co ( e^{(1)},\cdots , e^{(D-1)})$ in a gap-representation.  If  $t=0$ then $\rho=\tau$ which is separable. If $0 < t<1$, let $\sigma = \widehat{\rho}^{(D-1)}$ then $\sigma \succ \phi$ and thus $t\cdot \sigma+(1-t)\cdot \tau \succ t \cdot \phi + (1-t)\cdot \tau =\rho$, which implies
\[f_{\phi  }( t)= F(t \cdot \phi +(1-t)\cdot \tau )=F( \rho ) \leq C_L^-\]
\[  = F (L\cdot \sigma +(1-L)\cdot \tau )\leq F (L\cdot \phi +(1-L)\cdot \tau )=f_{\phi}( L) \;.\]
 The Lemma  then implies that $t \leq L$ and thus $\rho=\phi_t$
is separable.\\

We now show that ${\mathbb K} = {\mathbb L}_{C_F}$. By the definitions of ${\mathbb K}$ and $C_F$, ${\mathbb K}\subseteq {\mathbb L}_{C_F}$ and since ${\mathbb L}_{C_F}$ is closed, the closure $\overline{\mathbb K}$ of ${\mathbb K}$  is contained in ${\mathbb L}_{C_F}$.  Also, ${\mathbb K} \subseteq {\cal S}^{sep}$, and since ${\cal S}^{sep}$ is closed, $\overline{\mathbb K}\subseteq {\cal S}^{sep}$. The claim is proved by showing that ${\mathbb L}_{C_F}\subseteq \overline{\mathbb K}$.  Suppose $\tau \neq \rho \in {\mathbb L}_{C_F}$. Then for every  $0\leq t < 1$, $F( t.\rho +(1-t).\tau)= f_{\rho} (t) < F( \rho)\leq C_F$ by the Lemma, so for such $t$'s $t.\rho +(1-t).\tau \in {\mathbb K}$. Take an increasing sequence $\{t_n \}$ in $[0,1)$ with $\lim_{n\to \infty} t_n =1$. Then $\rho_n = t_n .\rho +(1-t_n).\tau \in {\mathbb K}$ and $\lim_{n\to \infty} \rho_n =\rho$ so that $\rho \in \overline{\mathbb K}$. This proves that $C_F$ satisfies the required properties and $C_F \leq C_L^+$. This completes the proof of the Theorem.\\

We will see that the bound $C_L^-$ is very poor and needs to be substantially improved in order to pin down $C_F$ (c.f., \S IV). This requires detailed information about the least separability modulus of the states $\widehat{\rho}^{(j)}$ as one parcours the maximal families of pairwise orthogonal pure states.\\

Theorem \ref{ULALA} states that the critical value of the $(D-1)$-th partial eigenvalue-sum is $1- (1-L)/D$.
Let us denote the critical value of the $k$-th partial eigenvalue-sum $\Sigma_k$ function by $C[k]$. The computation of $C_L^{\pm}$ for the $k$-th partial-sums is inmediate, and one gets
\begin{propo} For $k=1,2, \cdots , D-1$,
$ k \left( \frac{L}{D-1}+\frac{1-L}{D} \right) \leq C[k] \leq k \; \frac{1-L}{D} \; + L $.\\
\end{propo}

Notice that the bounds coalesce for $k=D-1$ to $C[D-1]=1-(1-L)/D$ recovering Theorem \ref{ULALA}.\\

The unitarily invariant strictly convex function $F( \rho ) = tr (\rho^2)$, is among the simplest separability detectors as it does not require spectral information to be calculated.   $C_L^{\pm}$ can be easily computed leading to:

\begin{propo} \label{spurcrit} The critical value $C_F$ for the trace of the square satisfies
$ \frac{D-1+L^2}{D(D-1)}  \leq C_F \leq \frac{L^2(D-1)+1}{D}$.\\

\end{propo}

\dem  For any $\rho$, we have $(t\cdot \rho+(1-t)\cdot \tau)^2= t^2\cdot \rho^2+2t(1-t)\cdot \rho\tau+(1-t)^2\tau^2= t^2\cdot \rho^2+(2t(1-t)/D)\cdot \rho+((1-t)^2/D)\tau$; hence $F( t\cdot \rho+(1-t)\cdot \tau)= t^2F( \rho )+ (1-t^2)/D$.
 If $\rho$ is pure $F(\rho )=1$ and thus $C_L^+= L^2+(1-L^2)/D$.
  If $spec ( \rho )=e^{(D-1)}$ and $F(\rho )=(D-1)^{-1}$; whence $C_L^-=F( L\cdot \rho +(1-L)\cdot \tau)= (L^2/(D-1))+ (1-L^2)/D$.$\Box$\\

Similar calculations can be carried out for other $F's$, e.g., the von Neumann entropy.

\section{Bipartite systems}

 For $N=2$ detailed entanglement information is available or obtainable. Notably, the results of Vidal and Tarrach, \cite{ViTa}, imply\footnote{\.{Z}yczkowski et al., obtain $L\geq 2/(2+D)$, \cite {Zy}.}
 \begin{equation} L = 2/(2+D) \;.\\ \label{L}\end{equation}

We can thus specify the characteristic parameters and bounds entering the results obtained in \S III.

\begin{propo} \label{biparcrit} For $N=2$, one has:
\begin{enumerate}
\item  $k(D+1)/(D+2)(D-1) \leq C[k] \leq (k+2)/(D+2)$ for $k=1,2, \cdots, D-2$ and
$C[D-1] = (D+1)/(D+2)$.
\item The critical value $C_F$ for $F(\rho)=tr (\rho^2)$ satisfies
\[ \frac{D (D+3)}{(D-1)(2+D)^2}  \leq C_F \leq \frac{D+8}{(2+D)^2}\;.  \]
\item The critical value $C_S$ of the von Neumann entropy satisfies
\[  \ln (2+D) - \frac{3}{2+D}\ln (3) \leq C_S \leq \ln ( D+2) - \frac{D+1}{D+2} \ln \left( \frac{D+1}{D-1}\right) \;.\]
\end{enumerate}
\end{propo}

\subsection{Qubit/qubit \& qubit/qutrit systems}

Consider the particular bipartite system where the first component is a qubit ($d_1=2$) and the second component is either a qubit ($d_2=2$) or a qutrit ($d_2=3$). By Eq. (\ref{L}), we have $L=1/3$ for the qubit/qubit system and $L=1/4$ for the qubit/qutrit system. \\

By a very elegant geometric method \.{Z}yczkowski et al., \cite{Zy},  obtained
\begin{propo} \label{ZY} If $N=2$ and $tr ( \rho^2) \leq 1/(D-1)$ then $\rho$ has positive partial transpose.\\
\end{propo}
Horodecki, Horodecki and Horodecki, \cite{HHH}, have shown  that  the positivity of the partial transpose is not only necessary but sufficient for separability in qubit/qubit and qubit/qutrit systems. Thus, since $tr( \rho^2) =1/(D-1)$ if $spec ( \rho ) = e^{(D-1)}$, the following result is a corollary to the result of \.{Z}yczkowski et al. (Proposition \ref{ZY}).
\begin{propo} \label{SEP} Suppose $N=2$, $d_1=2$ and $d_2=2,3$. If $spec ( \rho )= e^{(D-1)}$, then $\rho$ is separable.
\end{propo}
An independent proof is obtained from the result in  Appendix A, where it is shown directly, i.e., not via Proposition \ref{ZY},  that $ spec ( \rho ) =e^{(D-1)}$ implies that $\rho$ has positive partial transpose. $\Box$\\

This new bit of separability information inmediately leads to the following improvement on the lower bound for the critical
value of Theorem \ref{CRIT}:
\begin{propo}\label{criti} If $F$ is either a unitarily invariant, strictly convex continuous function or one of $\Sigma_k ( \cdot )$ for $k=1,2, \cdots, D-1$, on the states of a qubit/qubit or  a qubit/qutrit systems then
\[ c_F \geq \inf_{t\in [0,1]} \{ F (t\cdot \sigma + (1-t)L\cdot\omega + (1-t)(1-L)\cdot  \tau \} \;,\]
where  $spec ( \sigma )=e^{(D-1)}$, $spec (\omega )=e^{(D-2)}$ with $\omega \sigma =\omega/(D-1)$.
\end{propo}
This is proved in Appendix B. \\

The computation of the above infimum is quite straightforward. With the lower bound of \.{Z}yczkowski et al., for the trace of the square\footnote{Incidentally the lower bounds of Proposition \ref{criti} for the trace of the square are: $10/36$ for two qubits, and $33/192$ for the qubit/qutrit system; well below the (exact) $1/3$ and $1/5$ of \.{Z}yczkowski et al.}, we get:
\begin{propo} For $F( \rho )=tr ( \rho^2)$, and $S(\rho )=-tr( \rho \ln ( \rho ))$ one has:
\begin{enumerate}
\item For two qubits $ C_F= 1/3$, $C[3]=5/6$, $C[2]=2/3$, $1/3 \leq C[1]\leq 1/2$ and  $1.034 \approx \ln ( 6/\sqrt{3}) \leq C_S \leq \ln (6)- (5/6)\ln (5/3)\approx 1.329$.
\item For a qubit/qutrit system, $1/5 \leq C_F \leq 7/32$, $C[5]=7/8$, $C[4]=3/4$, $9/16\leq C[3]\leq 5/8$, $3/8\leq C[2] \leq 1/2$, $3/16\leq C[1]\leq 3/8$, and $1.667 \approx 3\ln (2)-(3/8)\ln (3)\leq C_S\leq 3\ln (2)-(3/4)\ln (3/2) \approx 1.775$.
\end{enumerate}
\end{propo}
The lower bound $2/3$ for $C[2]$ in two qubits also follows from Proposition \ref{BESSER} of Appendix B.\\

For a qubit/qutrit system we can now improve Theorem \ref{ULALA} as follows:
\begin{teor} \label{HOPLA} For a qubit/qutrit systems, if $spec ( \rho )=(\lambda_1, \cdots , \lambda_6)$ and $3 \lambda_6+ 5 \lambda_5 \geq 1$ then $\rho $ is separable. For each $s\in [0,1)$ there are entangled states with $3 \lambda_6+ 5 \lambda_5 =s$.
\end{teor}
This is a consequence of Proposition \ref{BESSER} of Appendix B,  Proposition \ref{SEP} and $L=1/4$.
Notice that the map $G$ defined on states of a qubit/qutrit system via $spec ( \rho )=(\lambda_1, \cdots, \lambda_5, \lambda_6)$ by $G(\rho)=3 \lambda_6 +5 \lambda_5 = 3+2 \Sigma_5 (\rho )-5 \Sigma_4 ( \rho )$ is unitarily invariant and continuous, but is not expected to be convex or concave. \\

\section{Final comments}
 With no other specific entanglement information other than $1>L>0$ we have obtained conditions on the spectrum of a state which guarantee its separability. These conditions are either direct restrictions on the spectrum or are hidden in the critical level-set of unitarily invariant,  convex, good continuous functions. Only the spectrum is needed to compute the values of such functions. We have also exemplified how more detailed entanglement information leads to less restrictive spectral conditions.\\

The problem of improving the results, in particular the lower bounds on $C_F$, seems worthy of pursuit. Exact values of $L$ for $N>2$ are not available. Further progess would come if something precise can be said about the following.
It is not difficult to show that there are entangled states $\rho$ with $spec ( \rho )=e^{(2)}$ and we have shown here that $spec ( \rho )=e^{(D-1)}$ implies the separability of $\rho$ in two cases ($N=2$, $d_1$ and $d_2=2,3$). Where in the spectral chain $e^{(D-1)} \succ e^{(D-2)} \succ \cdots \succ e^{(2)}$ is the cut separable/entangled? The solution of this problem will enhance the use of the gap-representation to deal with entanglement problems.\\

For a fixed compositum, call ${\cal F}$ the collection of the unitarily invariant, convex, good continuous functions on the state space. Take an $F \in {\cal F}$ and apply the $F$-test: reject the state $\rho$ if $F(\rho )\leq c_F$. You are left with the states passing the test $\{ \rho : \; F( \rho )> C_F\}$; this includes all pure states, in particular the pure product-states.  Take another $G \in {\cal F}$ and do the $G$-test on $\{ \rho : \; F( \rho )> C_F\}$ which outputs $
\{ \rho : \; F( \rho )> C_F\}\cap \{ \rho : \; G( \rho )> C_G\}$. When you exhaust ${\cal F}$ you are left with $\cap_{F \in {\cal F}}\{ \rho :\; F( \rho ) > C_F \}$ which still contains all pure states, hence all pure separable states.
Does one have $\cap_{F \in {\cal F}}\{ \rho :\; F( \rho ) > C_F \}= \{\rho \mbox{ is entangled}\}\cup \{\rho \mbox{ is a pure separable state}\}$? To put it another way: If $\rho$ is separable but not pure, is there an $F \in {\cal F}$ such that $F(\rho )\leq C_F$?  This would give an alternative definition of entanglement since it is easy to filter away the pure separable states. It would also solve the problem of thermal equilibrium states beautifully: $\rho_{\beta}$ is entangled iff $\beta$ does not lie in $[\beta_c^-,\beta_c^+]$. \\

I should like to thank Oscar Nagel for discussions and Omar Osenda for suggesting a look at thermal states. \\

\appendix

\section{Proof of Proposition \ref{SEP}}

\begin{propo} If $N=2$ and $spec (\rho )=e^{(D-1)}$  then the partial transpose of $\rho$ with respect to any of the subsystems is positive.
\end{propo}

\dem  We may assume that $d_1\leq d_2$. There is a unit vector $\psi \in {\cal H}_D$ such that $\rho = (D-1)^{-1} ( {\bf 1}-|\psi\rangle \langle \psi |)$. Let $\{ \psi_j :\; j=1, \cdots , d_1\}$ be an orthonormal basis of ${\cal H}_{d_1}$ and $\{ \eta_j:\; j=1,2, \cdots , d_2\}$ be an orthonormal basis of ${\cal H}_{d_2}$ such that the Schmid Decomposition of $\psi$ is $\psi = \sum_{j=1}^{d_1} \sqrt{\lambda_j} (\psi_j \otimes \eta_j)$. One has
\[ |\psi\rangle \langle \psi | = \sum_{j,k=1}^{d_1} \sqrt{\lambda_j\lambda_k} \, |\psi_j \otimes \eta_j\rangle \langle \psi_k\otimes \eta_k  |\;,\]
and $tr( |\psi\rangle \langle \psi |)=1$ implies that $\sum_{j,k=1}^{d_1} \sqrt{\lambda_j\lambda_k}=1$ in particular
$\sqrt{\lambda_j \lambda_k } \leq 1$ for every $j,k \in \{1,2,\cdots , d_1\}$.
Now denote by $A^T$ the partial transpose of the operator $A$ with respect to ${\cal H}_{d_2}$; that is to say $A^T$ is the linear operator associated to the matrix
\[ \langle \psi_j\otimes \eta_k , A (\psi_{j'}\otimes \eta_{k'})\rangle =\langle \psi_j\otimes \eta_{k'} , A (\psi_{j'}\otimes \eta_k)\rangle \]
via the orthonormal basis $\{\psi_j\otimes \eta_k : \; j=1,2,\cdots , d_1\;;\;k=1,2, \cdots , d_2 \}$. Then
\[ |\psi\rangle \langle \psi |^T = \sum_{j,k=1}^{d_1} \sqrt{\lambda_j\lambda_k} \, |\psi_j \otimes \eta_k\rangle \langle \psi_k\otimes \eta_j  | \;.\]
It follows that the operator $({\bf 1} -|\psi\rangle \langle \psi |)^T= {\bf 1}-|\psi\rangle \langle \psi |^T $ decomposes into a (direct) sum
\[   = \sum_{j=1}^{d_1} (1-\lambda_j)|\psi_j \otimes \eta_j\rangle \langle \psi_j\otimes \eta_j  |
 + \sum_{j=1}^{d_1}\sum_{k=d_1+1}^{d_2} |\psi_j \otimes \eta_k\rangle \langle \psi_j\otimes \eta_k  |
 \]
 \[ + \sum_{1 \leq j < k \leq d_1} \left( |\psi_j \otimes \eta_k\rangle \langle \psi_j\otimes \eta_k  | +
 |\psi_k \otimes \eta_j\rangle \langle \psi_k\otimes \eta_j   |\right.\]
 \[\left. - \sqrt{\lambda_j\lambda_k} |\psi_j \otimes \eta_k\rangle \langle \psi_k\otimes \eta_j  |-\sqrt{\lambda_j\lambda_k} |\psi_k \otimes \eta_j\rangle \langle \psi_j\otimes \eta_k  |\right) \;.\]
The second sum, which is present for $d_2 > d_1$, is manifestly positive. So is the first sum since $\lambda_j \leq 1$. The last  sum is a direct sum of $d_1(d_1-1)/2$ summands each of which has the matrix form
\[ \left( \begin{array}{cc}
1& -\sqrt{\lambda_j\lambda_k}\\
-\sqrt{\lambda_j\lambda_k} &1 \end{array} \right) \;,\]
with eigenvalues $1\pm \sqrt{\lambda_j\lambda_k}$ which are non-negative.$\Box$

\section{Proof of Theorem \ref{HOPLA}, and Proposition \ref{criti}}

Both results below are proved under the Hypothesis that:
\begin{equation} \mbox{ if $spec ( \rho )= e^{(D-1)}$, then $\rho$ is separable.}\label{god} \end{equation}
This has been proved (Proposition \ref{SEP}) for $N=2$, $d_1=2$ and $d_2=2,3$.

\begin{propo} \label{BESSER} Assume (\ref{god}).  Then, if $spec ( \rho ) = (\lambda_1, \lambda_2, \cdots , \lambda_D)$ and
\begin{equation}  \left( 1 - \frac{(1-L)(D-1)}{D}\right) \lambda_D +  \frac{(1-L)(D-1)}{D}\lambda_{D-1}  \geq (1-L)/D \;;\label{schranke}\end{equation}
it follows that $\rho$ is separable. Moreover, for every $s \in [0, (1-L)/D )$ there is an entangled state $\rho$ where the left-hand side of (\ref{schranke}) is equal to $s$.\end{propo}

Before proving this, we observe that the left-hand side of inequality (\ref{schranke}) satisfies ($\lambda_D \leq \lambda_j$)
\[ \lambda_{D-1}\geq \left( 1 - \frac{(1-L)(D-1)}{D}\right) \lambda_D + \frac{(1-L)(D-1)}{D} \lambda_{D-1}\geq \lambda_D \;,\ \]
so we  recover   Theorem \ref{ULALA}.\\

\dem Take any gap-representation $\rho=\sum_{j=1}^D t_j \widehat{\rho}^{(j)}$. Let $t=t_{D-1}$, and $s=\sum_{j=1}^{D-2}t_j$; then $t+s=1-t_D$\\
If $t=0$ then 
\begin{equation} \rho = \sum_{j=1}^{D-2}t_j\cdot \widehat{\rho}^{(j)}+ (1-s)\cdot \tau = s\cdot \omega + (1-s)\cdot \tau \;,\label{t=0}
\end{equation}
where
\[ \omega = \sum_{j=1}^{D-2}\frac{t_j}{s} \cdot \widehat{\rho}^{(j)} \;,\;\; \mbox{ if $s>0$}\;.\]
If $1>t>0$, then 
\begin{equation} \rho = t\cdot \sigma + (1-t)\cdot ( r \cdot \omega +(1-r)\cdot \tau ) \;\label{t>0}\end{equation}
where: $r= s/(1-t)$, $\omega $ is as above when $s>0$,
and   
\begin{equation}  \sigma =  \widehat{\rho}^{(D-1)} \label{sigma}
\end{equation}
is separable by (\ref{god}).
If $t=1$, then $\rho=\sigma$ is separable.   \\

If now $s\leq L (1-t)$, we have the following alternatives:
(1) $t=0$ and $s\leq L$, in which case Eq. (\ref{t=0}) implies that $\rho$ is separable;
(2) $1>t>0$ and $r \leq L$ in which case Eq. (\ref{t>0}) implies that $\rho$ is separable; and (3) $t=1$ in which case $\rho=\sigma$ is separable. Thus $s\leq L(1-t)$ implies that $\rho$ is separable.\\

 But since $s=1-t-t_D$ this inequality is equivalent to $(1-L)(1-t) \leq t_D$. In the notation of Proposition \ref{GR}, $t_j=\mu_j ( \lambda)$ is given by Eq. (\ref{gapcoeff}). Thus, we get $(1-L)(1-(D-1)(\lambda_{D-1}-\lambda_D)\leq D\lambda_D$ which is (\ref{schranke}).\\

 Take a maximally entangled state $\phi$ which is pure and suppose $0 \leq s < (1-L)/D$, then $1\geq  1-sD>L$ and the state $\phi_{1-sD}=(1-sD)\cdot \phi + sD \cdot \tau $ is entangled and has  $spec ( \phi_{1-sD})=(1-sD+s,s,s, \cdots, s)$;  we see that the left-hand side of (\ref{schranke}) is equal to $s$.$\Box$\\

The improvement on the lower bound for the critical value of a unitarily invariant, convex, good  continuous function, is the following:
\begin{propo} Assume (\ref{god}). If $F$ is either a unitarily invariant  strictly convex, continuous function, or one of the $\Sigma_k ( \cdot )$ with $k=1,2,\cdots, D-1$, then its critical value is not  below the number
\[  \inf_{t\in [0,1]} \{ F (t\cdot \sigma+(1-t) L\cdot \omega+(1-L)(1-t)\cdot \tau )\}\;,\]
where  $spec ( \omega )=e^{(D-2)}$, $spec ( \sigma )=e^{(D-1)}$, with $ \omega \sigma = \omega /(D-1)$.
\end{propo}

\dem  Consider a unitarily invariant, convex  continuous function $F$. Refer to the previous proof, whose notation $t,s,r, \omega, \sigma$ we keep. We have
\[ \rho = t\cdot \sigma +(1-t)\cdot (r\cdot \omega +(1-r)\cdot \tau)=
t\cdot \sigma +s\cdot  \omega +(1-t-s)\cdot \tau)\]
with $s=r(1-t)\leq (1-t)$. Notice that $spec ( \omega ) \in co (e^{(1)}, \cdots , e^{(D-2)})$.\\

 The first observation is that $t'\cdot \sigma +(1-t') r'\cdot \widehat{\rho}^{(D-2)}+(1-r')(1-t')\cdot \tau\succ t'\cdot \sigma+(1-t') r'\cdot \omega+(1-r')(1-t')\cdot \tau  $ for all $t',r'\in [0,1]$; indeed this is equivalent to $(1-t')r'(\Sigma_k (\omega ) -\Sigma_k(\widehat{\rho}^{(D-2)} )) \geq 0$ for $k=1,2,\cdots , D$; and this is satisfied because  $\omega \succ \widehat{\rho}^{(D-2)}$.\\

The other basic observation here is:
\begin{lema}  Suppose $F$ satisfies the hypothesis of the proposition and $\; 0 \leq  t' <1$.  Then the map $[0, 1-t']\ni s' \mapsto f_{\omega , \sigma} (s';t')= F( s'\cdot \omega+t'\cdot \sigma +(1-t'-s')\cdot \tau )$ is increasing and  convex. \label{>>}
\end{lema}

\dem    If $s_1,s_2 \in [0, 1-t']$ and $u \in [0,1]$ then $us_1+(1-u)s_2 \in [0, 1-t']$ and
\[ (us_1+(1-u)s_2)\cdot \omega + t'\cdot \sigma +(1-(us_1+(1-u)s_2)-t')\cdot \tau
\]
\[ =  u \cdot ( s_1.\cdot \omega + t'\cdot \sigma +(1-s_1-t')\cdot \tau ) \]
\[ +(1-u) \cdot (  s_2.\cdot \omega + t'\cdot \sigma +(1-s_2-t')\cdot \tau )\]
so that
\[ f_{\omega , \sigma }(us_1+(1-u)s_2 ;t') \]
\[ = F ((us_1+(1-u)s_2)\cdot \omega + t'\cdot \sigma +(1-(us_1+(1-u)s_2)-t')\cdot \tau ) \]
\[ = F(u \cdot ( s_1\cdot \omega + t'\cdot \sigma +(1-s_1-t')\cdot \tau ) \]
\[ +(1-u) \cdot (  s_2\cdot \omega + t'\cdot \sigma +(1-s_2-t')\cdot \tau )) \]
\[ \leq u F(s_1\cdot \omega + t'\cdot \sigma +(1-s_1-t')\cdot \tau ) \]
\[ +(1-u)F(s_2\cdot \omega + t'\cdot \sigma +(1-s_2-t')\cdot \tau ) \]
\[ = uf_{\omega , \sigma } (s_1;t')+(1-u)f_{\omega , \sigma }(s_2;t')\;;\]
so $f_{\omega , \sigma} (\cdot ;t')$ is convex. Moreover, if $F$ is strictly convex then the inequality is strict for $0<u<1$ and $s_1\neq s_2$ by Proposition \ref{F}, since $s_1\cdot \omega + t'\cdot \sigma +(1-s_1-t')\cdot \tau \neq s_2\cdot \omega + t'\cdot \sigma +(1-s_1-t')\cdot \tau$.\\

We  now  prove that $f_{\omega , \sigma }( \cdot ;t' )$ is increasing. If $0 \leq s_1 < s_2\leq (1-t')$, then
$ s_1\cdot \omega +t'\cdot \sigma +(1-s_1-t')\cdot \tau \succ s_2\cdot \omega + t'\cdot \sigma +(1-s_2-t')\cdot \tau $ because
the partial sums $\Sigma_k$ satisfy ($k=1,2,\cdots, D$)
\[   \Sigma_k (s'\cdot \omega + t'\cdot \sigma +(1-t'-s')\cdot \tau) \]
\[ = s'\Sigma_k ( \omega )+t'\Sigma_k ( \sigma )+(1-t'-s') \Sigma_k ( \tau )\;,\]
and
\[ \Sigma_k (s_1\cdot \omega + t'\cdot \sigma +(1-s_1-t')\cdot \tau) \leq  \Sigma_k (s_2\cdot \omega+ t'\cdot \sigma +(1-s_2-t')\cdot \tau )\]
is equivalent to
\[  (s_2-s_1) \Sigma_k ( \omega ) \geq (s_2-s_1) \Sigma_k ( \tau ) \;,\]
which is always satisfied because $\tau \succ \rho$ for every state $\rho$. Eq. (\ref{MONOTONIE}), implies that $F(s_1\cdot \omega+ t\cdot \sigma +(1-s_1-t)\cdot \tau ) \leq F(s_2\cdot \omega + t'\cdot \sigma +(1-s_2-t')\cdot \tau)$, or $f_{\omega , \sigma }(s_1;t')\leq f_{\omega , \sigma }(s_2;t')$. But as a non-decreasing  convex function,
$f_{\omega , \sigma }(\cdot ;t')$  must be constant up to a certain $s_* \leq 1$ and  increasing for $s\geq s_*$.
We have
\[ s_* = \sup \{ s'\in [0,1]: \; f_{\omega , \sigma }(s';t')=f_{\omega , \sigma }(0,;t')\}\;.\]
If now $F$ is strictly convex then $s_*=0$. Suppose $F=\Sigma_k ( \cdot )$ for some $k=1,2, \cdots , D-1$.
Then $f_{\omega, \sigma }(s_*;t')=f_{\omega , \sigma }(0;t')$ is equivalent to $s_* ( \Sigma_k ( \omega )- k/D)=0$ and Lemma \ref{ABS} implies that $s_*=0$.$\Box$\\

Put $H$ for the infimum that is claimed to be a lower bound for $C_F$, and assume that $F( \rho ) \leq H$. If $t=1$ we have $\rho=\sigma$ which is separable.   Otherwise, by the definition of $H$ and the first observation,
\[ f_{\omega , \sigma}(s;t)=F( \rho ) \leq H \leq F( t\cdot \sigma +(1-t)L\cdot \widehat{\rho}^{(D-2)}+(1-t)(1-L)\cdot
\tau ) \]
\[ \leq F( t\cdot \sigma +(1-t)L\cdot \omega +(1-t)(1-L)\cdot
\tau )=f_{\omega , \sigma } (L(1-t);t) \;.\]
And, by the Lemma, $s\leq L(1-t)$ which implies $r \leq L$ and the separability of $\rho$ follows from Eq. (\ref{t=0}) or Eq. (\ref{t>0}). Thus $C_F \geq H$.  This proves the claim. \\

A detailed analysis of the proof above suggests the introduction of a condition on unitarily invariant convex functions which we propose in the next appendix.

\section{$k$-good functions}

Suppose $F$ is a unitarily invariant, convex continuous function. Take an arbitrary maximal family $\{ \rho^{(j)}:\; j=1,2,\cdots , d\}$ of pairwise orthogonal pure states. $F$ is determined by its values in $\{ \sum_{j=1}^d \lambda_j
\rho^{(j)} : \; \lambda \in {\cal L}_d\}$. Fix $k\in \{1,2,\cdots , d-1\}$ and suppose  $spec ( \omega ) \in co ( e^{(1)}, e^{(2)}, \cdots , e^{(d-k)})$. For $t_1,t_2,\cdots ,t_{k-1}  \in [0,1]$ with $t= t_1+t_2+ \cdots +t_{k-1} \leq 1$ consider $f(s)= F( s\cdot \omega +\sum_{j=1}^{k-1} t_j \cdot \widehat{\rho}^{(d-k+j)} + (1-t-s)\cdot \tau )$, for $s\in [0,1-t]$, which is convex. Now, if $1-t\geq s>s'\geq 0$, it follows that $s'\cdot \omega +\sum_{j=1}^{k-1} t_j \cdot \widehat{\rho}^{(d-k+j)} + (1-t-s')\cdot \tau \succ s\cdot \omega +\sum_{j=1}^{k-1} t_j \cdot \widehat{\rho}^{(d-k+j)} + (1-t-s)\cdot \tau$, and by Eq. (\ref{MONOTONIE}), $f$ is non-decreasing. We say $F$ is {\em $k$-good} if $f(s)>f(0)$ for $s>0$. By the convexity of $f$ this is equivalent to $f$ is increasing.

 The proof of the last Proposition of the previous appendix will in fact work if $F$ is $2$-good.

\begin{propo} If $F$ is a unitarily invariant, convex , continuous function then it is good iff it is $1$-good. If $F$ is unitarily invariant, continuous and either strictly convex or one of the $\Sigma_k ( \cdot )$ for some $k=1,2,\cdots , d-1$, then $F$ is $p$-good for every $p=1,2,\cdots ,d-1$.
\end{propo}

\dem $1$-goodness is $F(s\cdot \rho +(1-s)\cdot \tau ) > F (\tau )$ for $s>0$ for every $\rho$ with $spec ( \rho )\in co ( e^{(1)},\cdots , e^{(d-1)})$. This property follows iff $F$ is good.\\

No matter what $p$ is, if $F$ is strictly convex, the map $f$ will be strictly convex and it then follows that it is increasing.\\

If $F=\Sigma_k $ for some $k=1,2, \cdots , d-1$. Then, for any $p=1,2,\cdots , d-1$,  $f (s)= f(0)$ is equivalent to $s (\Sigma_k ( \omega )- \Sigma_k ( \tau ))=0$ for $spec ( \omega ) \in co (e^{(1)}, \cdots, e^{(d-p)})$; which implies $s=0$ by Lemma \ref{ABS}.$\Box$\\

 \end{document}